\documentclass[aps,prf,twocolumn,amsmath,amssymb,amsfonts]{revtex4-2}
\usepackage{dcolumn}
\usepackage{bm}
\usepackage{amsmath}
\usepackage{graphicx} 
\usepackage{epstopdf}

\usepackage{xcolor}
\newcommand{\be}{\begin{equation}}
\newcommand{\ee}{\end{equation}}
\newcommand{\bea}{\begin{eqnarray}}
\newcommand{\eea}{\end{eqnarray}}
\newcommand{\rIm}{\mathrm{Im}}
\newcommand{\rRe}{\mathrm{Re}}
\newcommand{\re}{\mathrm{e}}            
\newcommand{\ri}{\mathrm{i}}
\usepackage[titletoc,title]{appendix}
\usepackage[normalem]{ulem} 

\begin{document}

\title{Excess shear force exerted on oscillating plate due to a nearby particle}

\author{Itzhak Fouxon$^{1,2}$}  \author{Boris Y. Rubinstein$^3$} \author{Alexander M. Leshansky$^1$}\email{lisha@technion.ac.il}
\affiliation{$^1$Department of Chemical Engineering, Technion, Haifa 32000, Israel}
\affiliation{$^2$Department of Computational Science and Engineering, Yonsei University, Seoul 120-749, South Korea}
\affiliation{$^3$ Stowers Institute for Medical Research, 1000 E 50th st.,Kansas City, MO 64110, USA}

\begin{abstract}

In the present paper we theoretically study the shear force exerted on an infinite horizontal plate undergoing fast lateral oscillations in presence of a rigid particle suspended in the viscous liquid above the plate. The study is largely motivated by Quartz Crystal Microbalance (QCM-D) technique which relies on analyzing response (complex impedance) of fast oscillating (in MHz range) quartz crystal disk in the liquid medium due to small substances \emph{adsorbed} at its surface. In fact, small substances \emph{suspended} in the liquid medium in the vicinity of the oscillating crystal may also contribute to impedance, as they modify the local shear force the suspending liquid exerts on the quartz crystal. For a dilute suspension the contributions of individual particles are additive and, therefore, our analysis is restricted to the excess shear force due a single spherical particle located at arbitrary distance above the plate. Three distinct cases are considered: (i) limiting case of high solid inertia, whereas the heavy particle can be considered as \emph{stationary}; (ii) a \emph{freely suspended} particle of arbitrary mass, undergoing fluid-mediated time-periodic rotation and translation and (iii) an \emph{adsorbed} particle moving with the plate as a whole without rotation. For small-amplitude plate oscillations the unsteady Stokes flow equations apply. We construct the series solution of these equations using the method of reflections, whereas its terms are written explicitly. Due to the exponential decay of the flow away from the oscillating plate, the truncated series containing only few low-order terms shows an excellent agreement with the rigorous numerical results for a wide range of particle sizes and  separation distances. The present results support the notion that the hydrodynamic contribution of the suspended small substances to the measured impedance is non-negligible or even dominant.

\end{abstract}

\maketitle

\section{Introduction \label{sec:intro}}

Quartz crystals are well known for their sharp resonance with the ratio of frequency to bandwidth (i.e., Q-factor), as high as $10^6$. This allows high precision measurements of small mass variance using quartz crystal microbalance (QCM) techniques \cite{review2015,lu,johans21}. The technique relies on the fact that small substances, e.g.,  colloidal particles, liposomes, macromolecules, viruses or bacteria adsorbed on the surface of the oscillating crystal, change both the frequency of the oscillations and their decay rate (see, e.g., \cite{qcm-nano1,qcm-nano2,qcm-nano3,fd,gold,lipo,slava18,alc19}). In vacuum, the frequency shift due to adsorbed mass provided by the QCM can be readily related to the mass change via the seminal Sauerbrey equation \cite{sauerbrey59}. The QCM in vacuum allows extremely accurate measurements of mass changes down to nanograms \cite{lu,johans21}. On the other hand, quantitative interpretation of the QCM-D (where ``D" stands for dissipation monitoring) measurements in liquids, introduced some $40$ years ago \cite{nomura80,qcm-liquid}, faces difficulties due to complex hydrodynamics due to either suspended or adsorbed particles, that have not yet been fully resolved \cite{review2015,slava18}. The present work is devoted to developing rigorous theoretical results on hydrodynamic interaction between a horizontally oscillating plate and a rigid particle (either adsorbed or freely suspended) that can be useful for interpretation of the QCM-D measurements in liquids.

The so-called \emph{small load approximation}, which holds as long as the frequency shift is much smaller than the frequency itself, is central to the interpretation of QCM-D data. The approximation implies that the QCM-D measures an area-averaged periodic stress $\overline{\sigma}$, i.e. the net shear force $\mathcal{F}$ exerted on the surface of the oscillating quartz crystal divided by its surface area \cite{review2015}. The impedance $\mathcal{Z}$ is then found as the ratio of $\overline{\sigma}/v_c$, where $v_c$ is the velocity amplitude of the crystal oscillations. Here $\mathcal{F}$ and $v_c$ and, therefore, $\mathcal{Z}$ are all complex quantities characterized by amplitude and phase. In the framework of the small load approximation, the shift in oscillation frequency $\Delta f$ and the shift in  oscillations' decay rate, $\Delta \Gamma$,  are both related to $\mathcal{Z}$ as $\Delta f-\ri \Delta\Gamma=\ri f \mathcal{Z}/(\pi \mathcal{Z}_Q)$, where the crystal impedance $\mathcal{Z}_Q$ is a known quantity \cite{review2015}. The particles affect the impedance via short-range forces operative at contact (e.g., due to elasticity of the adhesive contact, inertia of the fluid surrounding the particle, etc.) and the long-range hydrodynamic forces.

\begin{figure}[t]
\includegraphics[width=0.75\columnwidth]{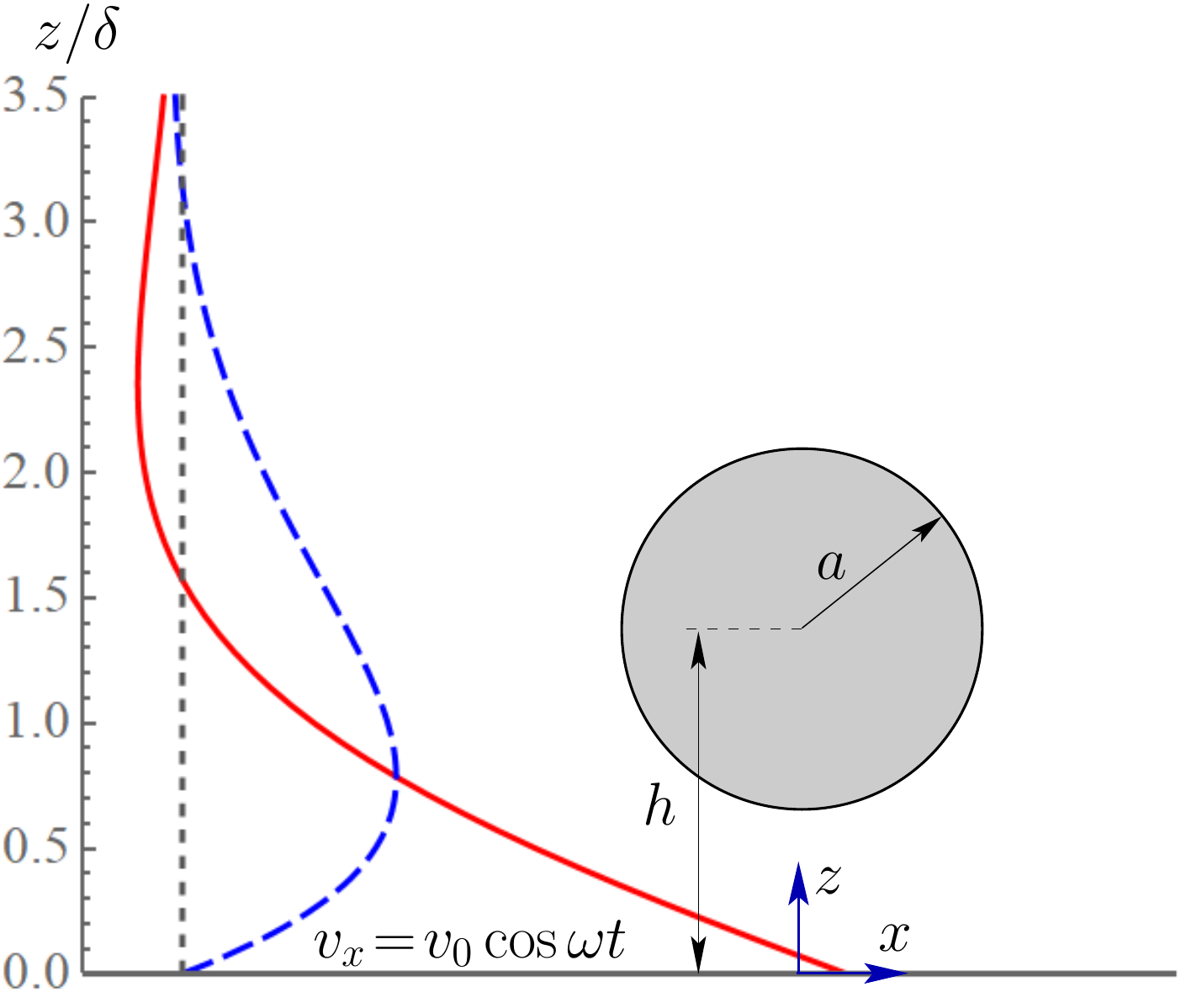}
\caption{Schematic illustration of the problem setup. A spherical particle of radius $a$ is positioned in the viscous liquid at distance $h$ above the horizontal plate at $z\!=\!0$ oscillating at MHz frequency with velocity $v_x=v_0 \cos{\omega t}$. The undisturbed (i.e., in the absence of the particle) Stokes velocity profiles, $v_x=v_0 \rRe[ \re^{-z/\delta} \re^{\ri(\omega t-z/\delta)}]$, are shown at $\omega t\!=\!0$ (solid, red) and $\omega t\!=\!\pi/2$ (dashed, blue) vs. the scaled vertical distance $z/\delta$. The short-dashed vertical line stands for the zero value of the velocity.
\label{fig:schematic}}
\end{figure}

In liquids, in contrast to vacuum, the shear force is applied to the oscillating crystal also in the absence of particles, due to viscous friction. It is well known (see, e.g., \cite{LL}), that the horizontal time-periodic oscillations of the infinite plane at $z\!=\!0$ with velocity $v_0\hat{\bm x} \cos{\omega t}$ create oscillatory unidirectional flow of the viscous liquid occupying the half-space $z\!>\!0$ with velocity given by the real part of $v_0 \hat{\bm x} \re^{-z/\delta} \re^{\ri(\omega t-z/\delta)}$ (see Fig.~\ref{fig:schematic}). Therefore, the disturbance created by the oscillating surface propagates upward as the transverse wave attenuated by the exponential factor with $\delta=\sqrt{2\nu/\omega}$, known as \emph{viscous penetration depth}. Here $\nu$ stands for the kinematic viscosity of the fluid. Notice that the above formula represents an exact solution satisfying the nonlinear Navier-Stokes equations and the no-slip boundary conditions at the surface of an oscillating plate. In presence of particles suspended in the fluid above an oscillating plate, no simple close-form solution is available even if the flow is approximated by the linear (unsteady) Stokes equations, derived from the full Navier-Stokes equations upon dropping the nonlinear inertia terms. The use of Stokes equations is justified provided that oscillation amplitude, $v_0/\omega$, is small in comparison to the particle size, $a$, which typically applies for QCM-D devices, where the oscillation amplitude is only about a few nanometers. The analysis could be further simplified by considering hydrodynamic interaction of a \emph{single} particle with the oscillating plate, assuming that the dispersion is dilute enough and that individual contributions of distinct particles to the net shear force are additive.

The \emph{excess} (i.e., due to presence of a particle) shear force exerted on the oscillating plate, $F$, depends on the particle radius, $a$, and the separation distance of its center from the plate, $h$ (see Fig.~\ref{fig:schematic}). The solution of the corresponding mathematical problem of determining $F$ is rather complex due to a non-trivial geometry and unsteadiness. Recently, this problem was considered theoretically in Ref.~\cite{Busca21} subject to a number of simplifying assumptions. Since the major focus of Ref.~\cite{Busca21} was the near-contact limit, $h\!\rightarrow\!a$, it has been assumed that the particle is adsorbed at the plate, implying that it undergoes in-sync oscillatory translations with the plate as a whole without rotation. Notice that no non-hydrodynamic forces were considered in \cite{Busca21} and the ``no slip" assumption at vanishing separations is of purely hydrodynamic origin. The authors refer to strong lubrication forces acting near contact and provide some numerical evidence to justify such assumption. To make analytical progress the authors made use of some extra simplifying \emph{ad hoc} assumptions and derived closed-form formulae for the hydrodynamic contribution to the excess shear force and impedance. The numerical computations performed for the oscillation frequency $f\!=\!35$~MHz (corresponding to $\delta\!=\!95$~nm, see also \cite{Busca2020}) and small particles with radii $a\!=\!15$--$100$~nm demonstrated a rather close agreement with the derived approximate formulae and it was argued that the theory should hold for $a/\delta \lesssim 2$. The major finding of \cite{Busca21} is that the contribution of hydrodynamic forces to the impedance, which is typically overlooked in interpretation of the QCM-D measurements \cite{review2015}, can, in fact, be substantial or even dominant. Large particles (with higher values of $a/\delta$) were not considered. Moreover, the linear velocity of the freely-suspended particles located at finite distance from the oscillating plate, is not equal (in magnitude and phase) to the plate velocity and they also free to rotate. The excess shear force, $F(h)$, due to a freely-suspended particle depends on its linear and angular velocity, which have to be determined self-consistently as part of the solution of the hydrodynamic problem, and it was not considered in \cite{Busca21}. These limitations, together with the unspecified range of validity of their analysis, call for further investigation.

In this paper we present a rigorous theoretical consideration of the excess shear force exerted by the viscous fluid on the horizontally oscillating infinite plane in presence of either heavy inertial (considered stationary), freely suspended or adsorbed (i.e, oscillating with the plate as a whole) particle of arbitrary size and mass. Analytical progress is possible using distant-particle approximation which relies on the solution of the unsteady Stokes equations by using the method of reflections. The translational and angular velocities of a freely suspended particle are determined self-consistently within the solution. The analytical results are then compared to the prediction of rigorous numerical computations performed using Finite Element Method (FEM). The limit of close proximity between the freely suspended particle and the oscillating plate is studied separately. The classical lubrication theory, which applies in this limit for steady Stokes equations \cite{kim}, was extended to unsteady flows. We provide the asymptotic conditions under which the lubrication approximation holds to transient flows and discuss validity of the ``no slip" condition of Ref.~\cite{Busca21}. The range of validity of the approximate theory in \cite{Busca21} is discussed in detail by comparing their formulae with the rigorous derivation via the boundary-integral formulation (in Appendix~\ref{app:A}).

We last remark that in a previous work \cite{fl18} we studied the dual problem of the force exerted on a particle oscillating in a fluid above a stationary infinite plane. Although the two problems are related, they do not reduce to each other.

\section{Shear force exerted on oscillating plate in presence of a particle}

The shear stress that the particle above the oscillating plane exerts on it depends on the particle's mobility. In this work we consider three different settings: (i) heavy inertial particle, which motion in response to the plate oscillations can be entirely neglected and which therefore considered stationary (see, e.g., \cite{prl20}), (ii) freely suspended particle of arbitrary mass, which oscillatory motion is not known in advance and has to be determined self-consistently as part of the solution, and (iii) adsorbed (i.e., adhered to the plate) particle which oscillates with the plate as a whole without rotation (see \cite{Busca21}).

\subsection{Stationary particle}\label{stationary}

We begin the analysis from the case (i) of a stationary spherical particle of radius $a$ with its center located at vertical distance $h$ above an infinite oscillating plate (see Fig.~\ref{fig:schematic}). This setting corresponds to the limit of high particle inertia, when the fluid-mediated particle's motion in response to the plate oscillations is too slow. The particle's inertia is characterized by the dimensionless Stokes number:
\begin{eqnarray}&&\!\!\!\!\!\!\!\!\!\!\!\!\!
\mathrm{St}\equiv \frac{m\omega}{\eta a}, \label{stokes}
\end{eqnarray}
where $m\!=\!4 \pi a^3 \rho_s/3$ is the particle's mass, $\omega$ is the frequency of plate oscillations and $\eta$ is the dynamic viscosity of the fluid. Up to a numerical factor, $\mathrm{St}$ is defined as the ratio of the relaxation (or Stokes) time $m/(6\pi \eta a)$ and the characteristic time scale of the flow, $1/\omega$.

The limit of stationary particle, described in this subsection, is the $\mathrm{St}\to\infty$ limit of the solution at arbitrary $\mathrm{St}$ to be derived later. This limit holds when either a particle is too heavy (or too large), or when the oscillations too fast, cf. \cite{prl20}. The analysis of the excess shear due to a stationary particle at $\mathrm{St}\gg 1$ is simpler than the general case of freely suspended particle corresponding to arbitrary $\mathrm{St}$. Moreover, as we shall see below, the solution of the problem (i) can be further used to find the excess shear force in other more complex settings (ii) and (iii).

We consider the setting where the fluid in the infinite half-space $z\!>\!0$ is set into motion by the periodic horizontal oscillations of the plate located at $z\!=\!0$ along the $x$-axis with frequency $\omega$ and amplitude $v_0$ (see Fig.~\ref{fig:schematic}). Assuming small amplitude of the oscillations, $v_0/\omega \ll a$, to the leading approximation the flow $\bm V$ is governed by the unsteady Stokes equations
\begin{eqnarray}&&\!\!\!\!\!\!\!\!\!\!\!\!\!
\partial_t \bm V\!=\!-\rho^{-1}\nabla P\!+\!\nu \nabla^2 \bm V,\ \ \nabla\cdot\bm V\!=\!0,
\nonumber\\&&\!\!\!\!\!\!\!\!\!\!\!\!\! \bm V(z=0)= v_0 \cos (\omega t)\hat{\bm x},\ \ \bm V(r=a, t)=0.\label{fud}
\end{eqnarray}
where $P$ is the pressure, $\rho$ and $\nu\!=\!\eta/\rho$ are the density and the kinematic viscosity of the fluid, respectively, and the radial distance $r=|\bm x-\bm x_c|$ is measured from the particle center located at $\bm x_c=(0, 0, h)$. We introduce dimensionless variables by normalizing fluid velocity with $v_0$, pressure with $\eta v_0/a$,  time with $\omega^{-1}$ and distance with $a$. Thus the dimensionless (complex) flow field $\bm v$ and pressure $p$ defined via $\bm V=v_0\rRe[\re^{-\ri \omega t} \bm v]$ and $P=\eta v_0\rRe[\re^{-\ri \omega t} p]/a$, where $\rRe$ stands for the real part, satisfy
\begin{eqnarray}&&\!\!\!\!\!\!\!\!\!\!\!\!\!
\lambda^2 \bm v\!=\!-\nabla p\!+\!\nabla^2\bm v,\ \ \nabla\cdot\bm v\!=\!0,
\nonumber\\&&\!\!\!\!\!\!\!\!\!\!\!\!\! \bm v(z=0)= \hat{\bm x},\ \ \bm v(r=1)=0. \label{fdon}
\end{eqnarray}
Here $\lambda^2\!=\!-\ri \mathrm{Ro}$ where $\mathrm{Ro}\!\equiv\!a^2\omega/\nu=2(a/\delta)^2$ is the Roshko number, characterizing the \emph{fluid} inertia. Notice that the Stokes number can be expressed as $\mathrm{St}\!=\!\mathrm{Ro}\xi$, where the parameter $\xi=m/\rho a^3$ (up to a factor of $4\pi/3$) corresponds to solid-to-fluid density ratio. Therefore, the solutions we derive below for different settings (i)--(iii) is written in terms of three dimensionless parameters: $a/\delta$, $\xi$ and $h/a$ (we could have alternatively used $\mathrm{Ro}$, $\mathrm{St}$ and $h/a$).

In the absence of a particle, the solution of Eqs.~(\ref{fdon}) is given by $\bm u_0=\re^{-\lambda z}\hat{\bm x}$, where $\lambda\!=\!(1-\ri)\sqrt{\mathrm{Ro}/2}\!=\!(1-\ri)\,(a/\delta)$, and $p_0\!=\!0$ (see, e.g. \cite{LL}). When the particle is present, no analytical solution of Eqs.~(\ref{fdon}) is readily available. The complication stems from the fact, that the Helmholtz equation, as opposed to the Laplace equation, is not separable in bi-spherical coordinates, for which $z\!=\!1$ and $r\!=\!1$ are coordinate surfaces. Some theoretical progress is however possible. The stress tensor corresponding to $\{\bm v, p\}$ in Eqs.~(\ref{fdon}) is defined by $\sigma_{ik}\!\equiv\! - p\delta_{ik}\!+\!\partial_k v_i\!+\!\partial_i v_k$. In absence of the particle this tensor has only $xz$ and $zx$ components, which at the plate equal to $-\lambda$. If the particle is present, this value of the stress at the plate holds far from the particle. Our major aim is determining the complex \emph{excess} shear force, $F_s$ exerted on the oscillating plate in the incompressible viscous liquid due to presence of the stationary particle located above the plate, defined by
\begin{eqnarray}&&\!\!\!\!\!\!\!\!\!\!\!\!\!
F_s\!=\!\int_{z=0}\! \left(\sigma_{xz}\!+\!\lambda\right) dx dy, \label{ff0}
\end{eqnarray}
where the subscript $s$ refers to the stationary particle. The solution decomposition $\bm v=\re^{-\lambda z}\hat{\bm x}+\bm u$, where $\bm u$ is the flow perturbation due to a stationary particle, yields:
\begin{eqnarray}&&\!\!\!\!\!\!\!\!\!\!\!\!\!
\lambda^2 \bm u\!=\!-\nabla p\!+\!\nabla^2\bm u,\ \ \nabla\cdot\bm u\!=\!0,
\nonumber\\&&\!\!\!\!\!\!\!\!\!\!\!\!\! \bm u(z=0)= 0,\ \ \bm u(r=1)=-\re^{-\lambda z}\hat{\bm x}. \label{fon17}
\end{eqnarray}
The stress tensor $\sigma_{ik}^u$ corresponding to $\{\bm u, p\}$ obeys
\begin{eqnarray}&&\!\!\!\!\!\!\!\!\!\!\!\!\!
\lambda^2 u_i\!=\!\frac{\partial \sigma^u_{ik}}{\partial x_k};\ \ \sigma_{ik}^u\!\equiv\!-p\delta_{ik}\!+\!\frac{\partial u_i}{\partial x_k}+\frac{\partial u_k}{\partial x_i}
\nonumber\\&&\!\!\!\!\!\!\!\!\!\!\!\!\!
=\sigma_{ik}+\left(\delta_{ix}\delta_{kz}+\delta_{iz}\delta_{kx}\right)\lambda \re^{-\lambda z},\label{st}
\end{eqnarray}
Thus the excess shear force $F_s$ can then be written as
\begin{eqnarray}&&\!\!\!\!\!\!\!\!\!\!\!\!\!
F_s\!=\!\int_{z=0} \sigma^u_{xz} dxdy=\!\int_{z=0} \partial_z u_{x}dxdy. \label{fofc}
\end{eqnarray}
The direct numerical study of the force using Eq.~(\ref{fofc}) could be problematic. The general structure of unsteady Stokes flows generated at the particle surface indicates that, at distances from the boundary greater than the viscous penetration depth $\delta/a\propto |\lambda|^{-1}$, the flow $\bm u$ a is a superposition of a potential (inviscid) flow and exponential correction, see, e.g. \cite{LL}. However the contribution of the dominant potential flow component into the integral in Eq.~(\ref{fofc}) vanishes identically. Hence $F_s$ is controlled entirely by the exponentially small correction to the potential flow. This renders accurate numerical computation of $F_s$ over infinite plate in Eqs.~(\ref{fofc}) challenging.

We rewrite $F_s$ in the form which is more suitable for the numerical study by using the Lorentz reciprocity \cite{kim}. For any dual flow satisfying $\lambda^2{\hat v}_i=\partial_k{\hat \sigma}_{ik}$ and the incompressibility condition we have:
\begin{eqnarray}&&\!\!\!\!\!\!\!\!\!\!\!\!\!
\frac{\partial ({\hat v}_i\sigma^u_{ik})}{\partial x_k}= \frac{\partial (u_i{\hat \sigma}_{ik})}{\partial x_k}. \label{lorentz}
\end{eqnarray}
Integrating Eq.~(\ref{lorentz}) over the volume of the semi-infinite domain and using the original flow field $\bm v$ satisfying Eqs.~(\ref{fdon}) as the dual flow, we find that:
\begin{eqnarray}&&\!\!\!\!\!\!\!\!\!\!\!\!\!
F_s\!=\!-\oint _{r=1}\!\!\re^{-\lambda z}\sigma_{xr}\,dS\!=\!-\oint _{r=1}\!\!\re^{-\lambda z}\sigma^u_{xr}\,dS
\nonumber\\&&\!\!\!\!\!\!\!\!\!\!\!\!\!
+\lambda \re^{-2\lambda h} \oint _{r=1} \re^{-2\lambda  \cos\theta} \cos{\theta}\, dS, \label{ifu}
\end{eqnarray}
where we made use of Eq.~(\ref{st}) and where $\theta$ stands for the spherical polar angle. The last integral in (\ref{ifu}) can be readily evaluated by using:
\begin{eqnarray}&&\!\!\!\!\!\!\!\!\!\!\!\!\!
\int _{0}^{\pi}\re^{-\zeta\cos{\theta}}\cos{\theta} \sin{\theta}\, d\theta=\int_{-1}^1 \re^{-\zeta x} x\, dx
\nonumber\\&&\!\!\!\!\!\!\!\!\!\!\!\!\!
=-2\frac{d}{d\zeta} \frac{\sinh\zeta}{\zeta}=2\frac{\sinh\zeta-\zeta\cosh\zeta}{\zeta^2}.
\end{eqnarray}

Thus, instead of integration over infinite plane at $z\!=\!0$ in Eq.~(\ref{fofc}), the excess shear force $F_s$ can be alternatively evaluated by integrating $\sigma^u_{xr}$ over the particle surface at $r\!=\!1$:
\begin{eqnarray}\!\!\!\!\!\!\!\!\!\!\!\!\!
F_s\!&=&\!-\oint_{r=1}\!\!\re^{-\lambda z} \sigma^u_{xr}dS
\nonumber\\\!\!\!\!\!\!\!\!\!\!\!\!\!
&& +\frac{\pi \re^{-2\lambda h} (\sinh{2\lambda}-2\lambda \cosh{2\lambda})}{\lambda}.\label{forcef}
\end{eqnarray}
Notice that the second (analytical) term in Eq.~(\ref{forcef}) is identical (up to a factor of $\pi$) to  expression for the ``viscous" contribution, $\hat{\mathcal{Z}}_\nu$, to the dimensionless impedance in Eq.~(13) of Ref.~\cite{Busca21}.

\subsection{Freely suspended particle}

We next consider the case (ii) corresponding to a rigid particle of an arbitrary size and mass suspended in the liquid above the oscillating plate (see Fig.~\ref{fig:schematic}) which is free to translate and rotate. Then instead of Eqs.~(\ref{fdon}), the flow is governed by the following set of equations (we use here the same notation for the fluid velocity, pressure and stress without ambiguity):
\begin{eqnarray}&&\!\!\!\!\!\!\!\!\!\!\!\!\!
\lambda^2 \bm v\!=\!-\nabla p\!+\!\nabla^2\bm v,\ \ \nabla\cdot\bm v\!=\!0,
\nonumber\\&&\!\!\!\!\!\!\!\!\!\!\!\!\! \bm v(z=0)= \hat{\bm x},\ \ \bm v(r=1)=V \hat{\bm x}+\Omega\, \hat{\bm y}\times \bm r. \label{fond}
\end{eqnarray}
Here $V\hat{\bm x}$ and $\Omega\hat{\bm y}$ stand for the unknown complex dimensionless translational and angular velocities of the particle, respectively. The angular velocity is scaled with $v_0/a$ and the real-valued (dimensionless) particle velocities are retrieved as $\rRe[V \re^{-\ri t}]$ and $\rRe[\Omega \re^{-\ri t}]$.  Notice that due to linearity of the unsteady Stokes equations, the plate oscillations result in particle displacement parallel to the plate (i.e., along the $x$-axis), and rotation about the $y$-axis (i.e., the same component as the vorticity generated in $xz$-plane in the absence of the particle).  The particle's translational and angular velocities are governed by the Newton's laws of motion that can be written as:
\begin{eqnarray}&&\!\!\!\!\!\!\!\!\!\!\!\!\!
\lambda^2 \xi V\!=\! \oint_{r=1}\!\! \sigma_{xr} dS,\nonumber\\&&\!\!\!\!\!\!\!\!\!\!\!\!\! \frac{2}{5} \lambda^2   \xi \Omega \!= \!\oint_{r=1}\!\! \left((z-h)\sigma_{xr}\!-\!x\sigma_{zr}\right) dS,\label{an}
\end{eqnarray}
where the parameter $\xi\!=\!m/(\rho a^3)$ was introduced before. Here we assumed that the particle has uniform density distribution with the corresponding moment of inertia.

\subsubsection*{Solution for the particle motion}
The excess stress exerted on oscillating plate due to a freely suspended particle depends on the particle's motion. Here we demonstrate how the particle's linear and angular velocities can be obtained using the solution of the auxiliary problem of a particle moving in the vicinity of a stationary plane. We apply the Reciprocal Theorem to the original flow $\bm v$ in Eq.~(\ref{fond}) using the dual solution $\{\widehat{\bm v}, \widehat{p}\}$ that obeys Eqs.~(\ref{fond}) under the boundary condition $\widehat{\bm v}\!=\!0$ at $z\!=\!0$, and $\widehat{\bm v}\!=\!\widehat{V} \hat{\bm x}+ \widehat{\Omega} \hat{\bm y}\times\bm r$ at $r\!=\!1$, where $\widehat{V}$ and $\widehat{\Omega}$ are some arbitrary linear and angular particle velocities.

Designating the stress tensor of the dual flow by $\widehat{\bm \sigma}$, we find that
\begin{eqnarray}&&\!\!\!\!\!\!\!\!\!\!\!\!\!
V\widehat{F}_x\!+\! \Omega\widehat{T}_y\!+\!\int_{z=0}\!\!\widehat{\sigma}_{xz} dxdy\!=\!\lambda^2\xi\left( \widehat{V} V\!+\! \frac{2}{5} \Omega \widehat{\Omega}\right), \label{fxty}
\end{eqnarray}
where we used Eqs.~(\ref{an}). The force and the torque that act on the sphere due to the dual flow are designated by $\widehat{\bm F}$ and $\widehat{\bm T}$, respectively. The force and the torque are linearly dependent on particle's translation and rotation velocities \cite{kim}:
\begin{eqnarray}&&\!\!\!\!\!\!\!\!\!\!\!\!\!
\left(\begin{array}{cc}
\widehat{F}_x\\
\widehat{T}_y  \end{array}\right)=\bm{\mathcal R} \left(\begin{array}{cc}
\widehat{V} \\
\widehat{\Omega}  \end{array}\right), \label{resist}
\end{eqnarray}
where $\bm{\mathcal R}$ is a symmetric resistance matrix \cite{note1}. We have by linearity $\int_{z=0}\widehat{\sigma}_{xz} dxdy={\mathcal A} \widehat{V}+{\mathcal B} \widehat{\Omega}$, where ${\mathcal A}$ and ${\mathcal B}$ are the resistance coefficients of the stationary plane due to oscillatory motion of a nearby rigid sphere. Substituting this ansatz together with (\ref{resist}) into (\ref{fxty}), we readily find that
\begin{eqnarray}&&\!\!\!\!\!\!\!\!\!\!\!\!\!
\left(\begin{array}{cc}
V \\
\Omega  \end{array}\right)^T\!\! \bm {\mathcal R}  \left(\begin{array}{cc}
\widehat{V} \\
\widehat{\Omega}  \end{array}\right)\!=\!-{\mathcal A} \widehat{V}\!-\!{\mathcal B} \widehat{\Omega}\!+\lambda^2\xi \left(V \widehat{V}\!+\! \frac{2 \Omega \widehat{\Omega}}{5}\right).
\end{eqnarray}
The above equation holds for any $\widehat{V}$ and $\widehat{\Omega}$ giving
\begin{eqnarray}&&\!\!\!\!\!\!\!\!\!\!\!\!\!
\left(\begin{array}{cc}
{\mathcal R}_{11}-\lambda^2 \xi   & {\mathcal R}_{12} \\
{\mathcal R}_{12} & {\mathcal R}_{22}-2 \lambda^2 \xi/5 \end{array}\right) \left(\begin{array}{cc}
 V \\
\Omega  \end{array}\right)=-\left(\begin{array}{cc}
{\mathcal A} \\
{\mathcal B} \end{array}\right).
\end{eqnarray}
Inverting the last equation one finds that
\begin{eqnarray}&&\!\!\!\!\!\!\!\!\!\!\!\!\!
\left(\begin{array}{cc}
 V \\
\Omega  \end{array}\right)=-\frac{1}{\left({\mathcal R}_{11}-\lambda^2 \xi \right)\left({\mathcal R}_{22}-2 \lambda^2 \xi/5 \right)-{\mathcal R}_{12}^2}
\nonumber\\&&\!\!\!\!\!\!\!\!\!\!\!\!\!
\times \left(\begin{array}{cc}
{\mathcal R}_{22}-2 \lambda^2 \xi/5   & -{\mathcal R}_{12} \\
-{\mathcal R}_{12} &  {\mathcal R}_{11}-\lambda^2 \xi \end{array}\right)\left(\begin{array}{cc}
{\mathcal A} \\
{\mathcal B} \end{array}\right).\label{motion}
\end{eqnarray}
Therefore, finding the velocities of the freely suspended inertial particle reduces to the calculation of the components of the resistance matrix $\bm{\mathcal R}$ and the resistance coefficients of the stationary plane, ${\mathcal A}$ and ${\mathcal B}$, to the oscillatory motion of a nearby rigid sphere.

\subsubsection*{Computation of the resistance coefficients}

In this subsection we obtain the resistance coefficients required for determining the velocity of the inertial particle freely suspended in the fluid above an oscillating plate in Eq.~(\ref{motion}). We first solve the auxiliary problem of a sphere translation parallel to the plate:
\begin{eqnarray}&&\!\!\!\!\!\!\!\!\!\!\!\!\!
\lambda^2 \bm v\!=\!-\nabla p\!+\!\nabla^2\bm v,\ \ \nabla\cdot\bm v\!=\!0,
\nonumber\\&&\!\!\!\!\!\!\!\!\!\!\!\!\! \bm v(z=0)=0,\ \ \bm v(r=1)=\hat{\bm x}.\label{flo}
\end{eqnarray}
The stress tensor of this problem $\bm \sigma^{t}$ defines the corresponding elements of the resistance matrix and $\mathcal A$:
\begin{eqnarray}
&& {\mathcal R}_{11}\!=\!\oint_{r=1}\!\! \sigma^t_{xr} dS\,, \nonumber \\
&& {\mathcal R}_{12}\!=\!{\mathcal R}_{21}\!=\!\oint_{r=1}\!\! \left((z-h)\sigma^t_{xr}\!-\!x\sigma^t_{zr}\right) dS\,, \nonumber \\
&& {\mathcal A}\!=\! \int_{z=0}\sigma^t_{xz} dxdy\,. \label{s}
\end{eqnarray}
To determine ${\mathcal R}_{22}$ and ${\mathcal B}$ we replace in Eqs.~(\ref{flo}) the boundary condition $\bm v(r=1)=\hat{\bm x}$ with $\bm v(r=1)=\hat{\bm y}\times \bm r$. Using the stress tensor  $\bm \sigma^{r}$ corresponding to the solution of the another auxiliary problem of a sphere rotation about the $y$-axis being parallel to the plate, we readily find:
\begin{eqnarray}&&\!\!\!\!\!\!\!\!\!\!\!\!\!
{\mathcal R}_{22}\!=\! \oint_{r=1}\!\! \left((z-h)\sigma^r_{xr}\!-\!x\sigma^r_{zr}\right) dS,\nonumber\\&&\!\!\!\!\!\!\!\!\!\!\!\!\!
{\mathcal B}=\int_{z=0}\sigma^r_{xz} dxdy.\label{t}
\end{eqnarray}
The particle's linear and angular velocities are then can readily be calculated using Eq.~(\ref{motion}).

The numerical computation of $\mathcal A$ and $\mathcal B$ by integration over the infinite surface of an oscillating plate at $z\!=\!0$, as given by Eqs.~(\ref{s})-(\ref{t}), can be challenging as was mentioned above. To avoid numerical inaccuracy, we rewrite them as integrals over the particle surface at $r\!=\!1$. Application of Eq.~(\ref{lorentz}) to the flows defined by Eqs.~(\ref{fdon}) and Eqs.~(\ref{flo}) gives
\begin{eqnarray}\!\!\!\!\!\!\!\!\!\!\!\!\!
{\mathcal A} \! &= & \! \oint _{r=1}\!\!\sigma_{xr}dS
\nonumber\\\!\!\!\!\!\!\!\!\!\!\!\!\!
&=& \oint _{r=1}\!\!\sigma^u_{xr}dS-\lambda \re^{-\lambda h}\oint _{r=1}\re^{-\lambda \cos{\theta}}\cos{\theta} dS =
\nonumber\\\!\!\!\!\!\!\!\!\!\!\!\!\!
&=& \oint _{r=1}\!\!\sigma^u_{xr}dS-\frac{4\pi (\sinh{\lambda}-\lambda \cosh{\lambda})}{\lambda}\, \re^{-\lambda h}\,, \label{os}
\end{eqnarray}
where $\bm \sigma^u$ is the stress tensor associated with the flow in Eqs.~(\ref{fon17}). Notice that ${\mathcal A}$ is equal to the force exerted by an oscillating plane on a fixed particle. The last (analytical) term in the RHS of Eq.~\ref{os} is identical (up to a factor $\pi$) to the ``kinetic" contribution, $\hat{\mathcal{Z}}_k$  to the dimensionless impedance in Eq.~(14) of Ref.~\cite{Busca21}.

A similar approach can be used to obtain the formula for ${\mathcal B}$. Application of Eq.~(\ref{lorentz}) to the flows defined by Eqs.~(\ref{fdon}) and Eqs.~(\ref{flo}) with the boundary condition $\bm v=\hat{\bm y}\times \bm r$ at $r\!=\!1$, yields:
\begin{eqnarray}
&& {\mathcal B}\!=\!\oint _{r=1}\!\!\epsilon_{iyk}r_k\sigma_{ir}dS\!=\!\oint _{r=1}\!\!\left(\cos{\theta} \sigma_{xr}\!-\!\sin{\theta} \cos{\phi}\, \sigma_{zr}\right)dS
\nonumber\\
&& =\!\oint _{r=1}\!\!\left(\cos{\theta}\,\sigma^u_{xr}\!-\!\sin{\theta} \cos{\phi}\, \sigma^u_{zr}\right)dS-\lambda \re^{-\lambda h} \mathcal{I}, \nonumber
\end{eqnarray}
where, as before, $\bm \sigma^u$ is the stress tensor of the flow in Eqs.~(\ref{fon17}) and $\mathcal{I}$ can be integrated analytically:
\begin{eqnarray}
\mathcal {I} &= &\oint _{r=1}\!\!\re^{-\lambda  \cos{\theta}} \left(\cos^2{\theta}\!-\!\sin^2{\theta} \cos^2{\phi}\right)dS \nonumber \\
&=& 4\pi \left(\frac{\sinh{\lambda}}{\lambda}- \frac{3\cosh{\lambda}}{\lambda^2} +\frac{3\sinh{\lambda}}{\lambda^3}\right). \nonumber
\end{eqnarray}
Finally, we have for ${\mathcal B}$:
\begin{eqnarray}
{\mathcal B} &=& \!\oint _{r=1}\!\!\left(\cos{\theta} \sigma^u_{xr}\!-\!\sin{\theta} \cos{\phi}\, \sigma^u_{zr}\right) dS
\nonumber\\
&&\! -4\pi \re^{-\lambda h}\left(\sinh{\lambda} +\frac{3\left(\sinh{\lambda}-\lambda\cosh{\lambda}\right)}{\lambda^2}\right).
\label{Bdefsph}
\end{eqnarray}

\subsubsection*{Excess shear force exerted on the plate}

After the particle velocities $V$ and $\Omega$ appearing in the boundary conditions in Eqs.~(\ref{fond}) are determined, one can compute the excess shear force exerted on the oscillating plate due to a freely suspended particle. The flow perturbation due to a freely suspended particle $\bm u=\bm v-\re^{-\lambda z}\hat{\bm x}$ is governed by:
\begin{eqnarray}&&
\lambda^2 \bm u\!=\!-\nabla p\!+\!\nabla^2\bm u,\ \ \nabla\cdot\bm u\!=\!0,\label{fon}
\\&& \bm u(z\!=\!0)=0,\ \ \bm u(r=1)=\left(V-\re^{-\lambda z}\right)\hat{\bm x}+ \Omega\hat{\bm y}\times \bm r,\nonumber
\end{eqnarray}
with the particle velocities $V$ and $\Omega$ are given by Eq.~(\ref{motion}). For $V\!=\!\Omega\!=\!0$ Eqs.~(\ref{fon}) reduce to Eqs.~(\ref{fon17}) describing flow around a stationary particle.  Thus, the solution of (\ref{fon}) can be obtained as superposition of the flow in Eq.~(\ref{fon17}), and the auxiliary flows due to translation and rotation of the sphere near the plate considered above [e.g., Eqs.~(\ref{flo})]. We thus conclude that the excess shear force the fluid exerts on the oscillating plane due to a freely suspended particle is given by
\begin{eqnarray}&&
F=\int_{z=0}\Sigma^u_{xz} dxdy=F_s+V {\mathcal A}+\Omega\, {\mathcal B}\,, \label{Fdef}
\end{eqnarray}
where $\bm \Sigma^u_{ik}$ is the stress tensor associated with the flow perturbation $\bm u$ in Eqs.~(\ref{fon}). Due to linearity of unsteady Stokes equations, the excess shear force is found as superposition of $F_s$ corresponding to a stationary particle and the respective contributions due to particle's translation and rotation.

In the limit of a heavy particle, $\mathrm{St}=|\lambda|^2\xi\!\to\!\infty$, one can expect that particle's motion becomes hindered by the solid inertia, and it readily follows from Eq.~(\ref{motion}) that:
\begin{eqnarray}&&\!\!\!\!\!\!\!\!\!\!\!\!\!
V\!=\!\frac{\mathcal A}{ \lambda^2 \xi}\!+\!o(\mathrm{St}^{-1})\,, \quad \Omega \!=\!\frac{5 \mathcal B}{2 \lambda^2 \xi}\!+\!o(\mathrm{St}^{-1})\,.
\label{VOm}
\end{eqnarray}
Substituting these asymptotic expressions into (\ref{Fdef}) we obtain the high-$\mathrm{St}$ asymptotic limit for the excess shear force exerted on the plate due to a freely suspended particle:
\begin{eqnarray}&&
F=F_s+\frac{{\mathcal A}^2}{ \lambda^2 \xi}+\frac{5 {\mathcal B}^2}{2 \lambda^2 \xi}\!+\!o(\mathrm{St}^{-1})\,. \label{Ffree}
\end{eqnarray}

\subsection{Adsorbed particle}

Let us now consider a case of an ``adsorbed" particle which firmly adheres to the plate and, therefore, oscillates in-sync with it as a whole \cite{Busca21}. Although from physical point of view particle adhesion corresponds to vanishing separation distance, $h\approx a$, we shall follow \cite{Busca21} and study the hypothetical setting of arbitrary proximity $h \geq a$. The flow around the particle oscillating in-sync with the plate satisfies
\begin{eqnarray}&&\!\!\!\!\!\!\!\!\!\!\!\!\!
\lambda^2 \bm v\!=\!-\nabla p\!+\!\nabla^2\bm v,\ \ \nabla\cdot\bm v\!=\!0, \nonumber\\
&&\!\!\!\!\!\!\!\!\!\!\!\!\! \bm v(z\!=\!0)= \hat{\bm x},\ \ \bm v(r\!=\!1)=\hat{\bm x}.
\end{eqnarray}
The perturbation of the flow, $\bm u=\bm v-\re^{-\lambda z}\hat{\bm x}$, is governed by:
\begin{eqnarray}&&\!\!\!\!\!\!\!\!\!\!\!\!\!
\lambda^2 \bm u\!=\!-\nabla p\!+\!\nabla^2\bm u,\ \ \nabla\cdot\bm u\!=\!0,
\nonumber\\&&\!\!\!\!\!\!\!\!\!\!\!\!\! \bm u(z\!=\!0)= 0,\ \ \bm u(r\!=\!1)=\left(1-\re^{-\lambda z}\right)\hat{\bm x}.
\end{eqnarray}
Using the superposition principle we find that the perturbation of the force the fluid exerts on the plane due to an adsorbed particle $F_a$ can be found as
\begin{eqnarray}&&\!\!\!\!\!\!\!\!\!\!\!\!\!
F_a=F_s+{\mathcal A}, \label{Fa}
\end{eqnarray}
where ${\mathcal A}$ is defined in Eqs.~(\ref{flo})-(\ref{s}). Using Eqs.~(\ref{forcef}) and (\ref{os}) this can be written as
\begin{eqnarray}\!\!\!\!
F_a\!&=&\!\oint_{r=1}\!\!\left(1\!-\!\re^{-\lambda z} \right)\sigma^u_{xr}dS\!-\!\frac{4\pi \re^{-\lambda h} (\sinh{\lambda}\!-\!\lambda \cosh{\lambda})}{\lambda}\,
\nonumber\\&&\!\!\!\!
 +\frac{\pi \re^{-2\lambda h} (\sinh{2\lambda}-2\lambda \cosh{2\lambda})}{\lambda}. \label{fa}
\end{eqnarray}
The last two (analytical) terms in the RHS comprise (up to a factor of $\pi$) the net hydrodynamic contribution to the complex impedance due to an adsorbed particle reported in Ref.~\cite{Busca21}. However, rigorous computation of the excess shear force due to a nearby (either stationary, freely suspended or adsorbed) particle, requires integration of the stress components over its surface at $r\!=\!1$ [see the integral terms in the expressions for $F_s$ and $F_a$ in Eqs.~(\ref{forcef}) and (\ref{fa}), respectively]. While such integration can be performed numerically (see Sec.~\ref{sec:num}), it is yet possible to make analytical progress in some asymptotic limits involving the three available length scales, $a$, $h$ and $\delta$, as we demonstrate in the following sections. The derivations are rigorous and do not involve any simplifications, apart from the assumption of small-amplitude oscillations that allowed to neglect the nonlinear inertia terms in the flow equations. The assumptions and validity range of the approximate theory in Ref.~\cite{Busca21} are examined in detail in Appendix~\ref{app:A}.

\section{Small-particle limit \label{sec:sma}}

The expressions above can be simplified in the small particle limit, $\delta \gg a$ or $|\lambda|\ll 1$. In the leading order the problem of determining excess shear force $F_s$ reduces to the problem of obtaining the force exerted on the particle oscillating with constant velocity near an infinite plane defined in Eq.~(\ref{flo}).

In this limit the velocity distribution over the particle surface in the last boundary condition in Eq.~(\ref{fon17}) is approximately constant, $\bm u(r\!=\!1)\approx -\re^{-\lambda h}\hat{\bm x}$ as readily follows from $\re^{-\lambda z}=\re^{-\lambda h-\lambda\cos\theta}\approx \re^{-\lambda h}$, where $\theta$ is the spherical angle of the radius vector from the particle center. The last approximation is also valid for $h\gg 1$, irrespective of $\lambda$. Thus,
\begin{eqnarray}&&\!\!\!\!\!\!\!\!\!\!\!\!\!
\bm u=-\re^{-\lambda h} \bm v+o\left(|\lambda|,\, \frac{1}{h}\right),
\end{eqnarray}
with $\bm v$ is defined in Eq.~(\ref{flo}) and the approximate equality holds provided that $|\lambda|^{-1}\gg 1$ or $h\gg 1$. Using $\re^{-\lambda z}\approx \re^{-\lambda h}$ in Eq.~(\ref{forcef}) we find
\begin{eqnarray}&&\!\!\!\!\!\!\!\!\!\!\!\!
F_s\!\approx \re^{-2\lambda h} {\mathcal R}_{11}+\!\frac{\pi \re^{-2\lambda h} (\sinh{2\lambda}\!-\!2\lambda \cosh{2\lambda})}{\lambda}.\label{Ffar}
\end{eqnarray}
Here ${\mathcal R}_{11}$ is given by the $x$-component of the force exerted on the particle that oscillates with unit velocity in $x$--direction [see in Eq.~(\ref{s})]. It has been  calculated in the limit $h\gg |\lambda|^{-1}\gg 1$ in \cite{fl18}:
\begin{eqnarray}&& \!\!\!\!\!\!\!\!\!\!\!\!\!
-{\mathcal R}_{11}=6\pi \left(1+\lambda+\frac{\lambda^2}{9}\right)
\nonumber\\&& \!\!\!\!\!\!\!\!\!\!\!\!\!
+\frac{3\pi}{2h^3}\left(1+\lambda+\frac{\lambda^2}{3}\right)\left[\frac{1}{6}+\frac{3}{4\lambda^2}\left(1+\lambda+\frac{\lambda^2}{9}\right)\right]. \label{R11}
\end{eqnarray}
In the other limiting case, $|\lambda|^{-1}\!\gg\! h\gg\! 1$, one can use (\ref{Ffar}) with ${\mathcal R}_{11}$ given by \cite{fl18}:
\begin{eqnarray}&&\!\!\!\!\!\!\!\!\!\!\!\!\!
-{\mathcal R}_{11}=6\pi \left(1+\frac{9}{16h}+\frac{9h\lambda^2}{8}\right). \label{R11a}
\end{eqnarray}
In both above limits (and also at $h\sim |\lambda|^{-1}\gg 1$) we have the Stokes result ${\mathcal R}_{11}\approx -6\pi$ to the leading approximation.  It further follows from Eq.~(\ref{Ffar}) that for $|\lambda|^{-1}\gg 1$ and $h\gg 1$ the excess shear stress due to a stationary particle reduces to
\begin{eqnarray}&&
F_s\approx\!-6\pi \re^{-2\lambda h}\,. \label{farfs}
\end{eqnarray}
Notice that for small particles with $|\lambda|^{-1}\gg 1$ at close proximity, $h\sim 1$, the Eq.~(\ref{farfs}) does not hold, however (\ref{Ffar}) with asymptotic expressions (\ref{R11}) or (\ref{R11a}) for ${\mathcal R}_{11}$ can be used instead.
For example, the prediction (\ref{Ffar}) with ${\mathcal R}_{11}$ given by (\ref{R11}) shown in Figs.~\ref{fig:Fs} by thick dashed lines, shows a close agreement with the results of the rigorous numerical solution (see Sec.~\ref{sec:num}) shown by solid lines. In the case of $a/\delta\!=0.25$ (or $|\lambda|^{-1}\approx 2.8$), the theory, developed for $h\gg |\lambda|^{-1}\gg 1$, holds reasonably well for $h\gtrsim |\lambda|^{-1}\gtrsim 1$ (roughly the small-particle prediction holds for separation distance down to $h\sim |\lambda|^{-1}$). For larger values of $a/\delta\gtrsim 1$, as might be anticipated, the prediction becomes inaccurate.

The small-particle approximation of the excess shear force $F_a$ can be found similarly. It follows from Eq.~(\ref{fa}) that
\begin{eqnarray}
F_a&\!=\!&\left(\re^{-2\lambda h}\!-\!\re^{-\lambda h} \right){\mathcal R}_{11}\!-\!\frac{4\pi \re^{-\lambda h}  (\sinh{\lambda}\!-\!\lambda \cosh{\lambda})}{\lambda} \, \nonumber\\
&& \!+\!\frac{\pi \re^{-2\lambda h}(\sinh{2\lambda}-2\lambda \cosh{2\lambda})}{\lambda} \!+\!o\left(|\lambda|, \frac{1}{h}\right)\,.\label{Fafar}
\end{eqnarray}
In the limit $|\lambda|^{-1}\gg 1$ and $h\gg 1$ we have
\begin{eqnarray}&&
F_a\approx\!6\pi\left(\re^{-\lambda h}\!-\!\re^{-2\lambda h} \right)\,. \label{farfa}
\end{eqnarray}
Thus for a small particle oscillating in-sync with the plane at a large distance, the force decays slower than for the stationary particle and has an opposite sign, cf. Eq.~(\ref{farfs}). The slower decay of $F_a$ with the distance could be anticipated, since the particle oscillating in-sync with the plane generates the flow in its vicinity, as opposed to the stationary particle which only perturbs the flow originated at the plate. This observation is also supported by the numerical results, see Fig.~\ref{fig:Fa}.

Finally, we consider the case of a freely suspended particle at $\delta \gg a$. We demonstrate that ${\mathcal A}$ and ${\mathcal B}$ can be written via ${\mathcal R}_{ik}$ by using Eqs.~(\ref{os}) and (\ref{Bdefsph}). Substituting $\bm u\approx-\re^{-\lambda h} \bm v$, as explained above before Eq.~(\ref{Ffar}), it can be shown that $\oint _{r=1}\!\!\sigma^u_{xr}dS$ in Eq.~(\ref{os}) equals to $-\re^{-\lambda h}{\mathcal R}_{11}$ at either $|\lambda|\ll 1$ or $h\gg 1$. Further using Eq.~(\ref{os}) yields
\begin{eqnarray}&&\!\!\!\!\!\!\!\!
{\mathcal A}\!= \!-\re^{-\lambda h}{\mathcal R}_{11}\!-\!\frac{4\pi\re^{-\lambda h} (\sinh{\lambda}\!-\!\lambda \cosh{\lambda})}{\lambda}
\!+\!o\left(|\lambda|, \frac{1}{h}\right). \nonumber 
\end{eqnarray}
When both $|\lambda|^{-1}\gg 1$ and $h\gg 1$, we have to the leading approximation
\begin{eqnarray}&&\!\!\!\!\!\!\!\!
{\mathcal A}\approx\!6\pi\re^{-\lambda h}\,, \label{farA}
\end{eqnarray}
Similarly we find that the first term in the RHS of Eq.~(\ref{Bdefsph}) equals to $-\re^{-\lambda h}{\mathcal R}_{12}$ giving
\begin{eqnarray}
{\mathcal B}&=&-\re^{-\lambda h}{\mathcal R}_{12}\!\nonumber \\
&& -4\pi\re^{-\lambda h}\frac{\lambda^2\sinh{\lambda}\!+\!3\left(\sinh{\lambda}\!-\!\lambda\cosh{\lambda}\right)}{\lambda^2}\!+\!o\left(|\lambda|, \frac{1}{h}\right).\nonumber
\end{eqnarray}
The viscous torque exerted on a sphere undergoing oscillatory translations in unbounded fluid vanishes, implying that $\lim_{h\to\infty} {\mathcal R}_{12}=0$. Thus, at large distances $|{\mathcal B}|\ll |{\mathcal A}|$ and in the limit $|\lambda|^{-1}\gg 1$ and $h\gg 1$, Eq.~(\ref{motion}) becomes
\begin{eqnarray}&&\!\!\!\!\!\!\!\!\!\!\!\!\!
\left(\begin{array}{cc}
 V \\
\Omega  \end{array}\right)=-\left(\begin{array}{cc}
{\mathcal R}_{11}^{-1} & 0\\
0 &  {\mathcal R}_{22}^{-1}\end{array}\right)\left(\begin{array}{cc}
6\pi\re^{-\lambda h} \\
0 \end{array}\right)=\left(\begin{array}{cc}
\re^{-\lambda h} \\
0 \end{array}\right). \label{farvl}
\end{eqnarray}
In derivation of (\ref{farvl}) we neglected the terms $\mathcal{O}(\lambda^2)$ and terms $\propto\!{\mathcal R}_{12}$ in Eq.~(\ref{motion}). It can be readily seen from Eq.~(\ref{farvl}) that a small particle away from the oscillating plate translates with the velocity of the undisturbed flow $\re^{-\lambda h}$ while its angular velocity vanishes.

As opposed to $F_s$ and $F_a$, the small-particle limit cannot be applied for finding the leading-order approximation for the excess shear force $F$ exerted on the plate due to a small freely suspended particle. This is due to the fact that the leading-order terms in
Eq.~(\ref{Fdef}) at $h\to\infty$ vanish. The excess shear force $F$ is determined by higher-order corrections and these are not readily available. On the other hand, the approximate distant-particle theory (see the next Section) provides very accurate prediction for the excess shear force for all three settings (stationary, absorbed and freely suspended particle) in a wide range of particle size, $a/\delta$. Thus, we shall not pursue further the asymptotic small-particle expansion, but instead focus on the distant-particle theory that relies on the method of reflections and provides remarkably accurate predictions for excess shear force even at close prosimity $h\!\gtrsim\!a$.

\section{Approximate theory for a distant particle \label{sec:dist}}

In this Section we present an approximate solution for the excess shear force exerted on the oscillating plate due to a distant particle suspended above it. We shall consider different settings (i.e., stationary, freely suspended and firmly attached particle) and examine the accuracy of the predictions by comparing them to the results of rigorous numerical computations. We remark that the results of the previous Sec.~\ref{sec:sma} indicate that in the limit of $h\gg 1$, irrespective of the particle size, the excess shear force reduces to the calculation of the resistance coefficients ${\mathcal R}_{ik}$. Therefore, the distant-particle limit could be obtained by deriving the corresponding higher order approximations for ${\mathcal R}_{ik}$. However, we find that constructing the flow perturbation due to a distance particle using the method of reflections is a more straightforward approach.

\subsection{Stationary particle}

We assume that $h\!\gg \mathrm{max}(a,\delta$), while the ratio $a/\delta$ is not constrained. Hereafter, unless told otherwise, the origin of the coordinates is assumed to be located at the center of the particle. Solution to Eqs.~(\ref{fon17}) for the flow perturbation due to a stationary particle can be obtained as a series of reflections  $\bm u=-\re^{-\lambda h}\bm u^0+\re^{-\lambda h}\bm u^1+\ldots$ where the dots stand for higher order terms \cite{hb}. The leading-order approximation satisfies
\begin{eqnarray}&&\!\!\!\!\!\!\!\!\!\!\!\!\!\!\!\!
\lambda^{2}\bm u^0\!=\!-\!\nabla p^0
\!+\! \nabla^2\bm u^0,\ \ \nabla\!\cdot\!\bm u^0\!=\!0,\nonumber\\&&\!\!\!\!\!\!\!\!\!\!\!\!\!\!\!\!
\bm u^0(r\!\rightarrow\! \infty)=0,\ \
\bm u^0(r\!=\!1)\!=\!\re^{-\lambda z}{\hat x}.\label{start}
\end{eqnarray}
The first-order correction obeys
\begin{eqnarray}&&
\lambda^{2}\bm u^1\!=\!-\!\nabla p^1\!+\! \nabla^2\bm u^1,\ \ \nabla\!\cdot\!\bm u^1\!=\!0,\nonumber\\&&
\bm u^1(z\!=\!-h)=\bm u^0(z\!=\!-h).\label{star}
\end{eqnarray}
The excess shear force defined in Eq.~(\ref{fofc}) is given, to the leading approximation, by
\begin{eqnarray}\!\!\!\!\!\!\!\!\!\!\!\!\!
F_s^0\! &=& \!-\re^{-\lambda h} \int_{z=-h} \partial_z u^0_{x} dxdy\nonumber\\\!\!\!\!\!\!\!\!\!\!\!\!\!\!\!\!
&=&\re^{-\lambda h} \frac{\partial}{\partial h}\int_{z=-h}  u^0_{x}(x, y, -h)  dxdy. \label{fo}
\end{eqnarray}
Thus the calculation of the force $F_s$ in this order demands finding the solution of unsteady Stokes equations in an infinite space with some prescribed velocity distribution at the surface of the unit sphere and the subsequent integration of the solution over the plane. However at this order the flow near the plate is not accurate as it does not satisfy the no-slip boundary conditions. As a result the contribution to the force exerted on a plate of $\bm u^1$ is of the same order as that given by Eq.~(\ref{fo}) and must be considered. The solution of Eq.~(\ref{star}) has zero pressure, $p^1=0$, and thus $\bm u^1$ in the semi-infinite domain is given by the Fourier transform:
\begin{eqnarray}&&\!\!\!\!\!\!\!\!\!
\bm u^1\!=\!\int\! \widetilde{\bm u}^0 (q_x, q_y)\,\re^{\ri q_x x+\ri q_y y- k(z+h)}\, \frac{dq_xdq_y}{(2\pi)^2},\\&&\!\!\!\!\!\!\!\!\!
\widetilde{\bm u}^0 (q_x, q_y)\!=\!\int \!\bm u^0(x, y, z=-h)\, \re^{-\ri q_x x-\ri q_y y} dxdy,\nonumber
\end{eqnarray}
where $k\equiv \sqrt{q^2\!+\!\lambda^2}$. By differentiating the above equation and integrating over the infinite plane gives the contribution to the shear force from $\bm u^1$:
\begin{eqnarray}&&\!\!\!\!\!\!\!\!\!\!\!\!\!
F_s^1\!=\!\re^{-\lambda h} \int_{z=-h} \partial_z u^1_{x} dxdy\nonumber\\&&\!\!\!\!\!\!\!\!\!\!\!\!\!\!\!\!
=-\lambda \re^{-\lambda h}  \int \!\bm u^0(x, y, z=-h)dxdy. \label{fo1}
\end{eqnarray}
This correction is of the same order as the one due to $\bm u^0$ in Eq.~(\ref{fo}). This is due to the fact that the correction decays exponentially fast away from the plane, however it is not small near the plane. Higher-order reflections yield negligible contributions and can be omitted.

Thus the calculation of the excess shear force in the limit of large separations reduces to the study of the unbounded transient Stokes flow $\bm u^0$ in Eqs.~(\ref{start}) driven by non-uniform velocity distribution at the boundary of the unit sphere. The general exterior solution of the unsteady Stokes problem was recently proposed in \cite{fouxon22}. We briefly review some relevant results in this Section. We study the general transient Stokes equations:
\begin{eqnarray}&&\!\!\!\!\!\!\!\!\!\!\!\!\!\!\!\!
\lambda^{2}\bm u\!=\!-\!\nabla p
\!+\! \nabla^2\bm u,\ \ \nabla\!\cdot\!\bm u\!=\!0, \label{is}
\end{eqnarray}
in polar spherical coordinates $(r,\theta,\phi)$. It is assumed that the pressure and velocity vanish at infinity. The general solution is given by superposition of three terms resembling the Lamb's decomposition of the solution of steady Stokes equations:
\begin{eqnarray}&&\!\!\!\!\!\!\!\!\!\!\!\!\!\!\!\!
\bm u=-\frac{\nabla p}{\lambda^2}+\nabla\times (\bm {\hat r} e^{\lambda (1-r)}X )+ \bm u^H, \label{su}
\end{eqnarray}
where $p$ is the pressure, $X$ is a scalar function and $\bm u^H$ is a solenoidal solution of the vector Helmholtz equation $ \nabla^2 \bm u^H=\lambda^2 \bm u^H$.

By taking divergence of Eqs.~(\ref{is}) it can be readily seen that pressure is a harmonic function. Therefore, it can be represented as,
\begin{eqnarray}&&\!\!\!\!\!\!\!\!\!\!\!\!\!
p=\sum_{l=1}^{\infty}\sum_{m=-l}^{m=l}\frac{c_{lm}Y_{lm}(\theta, \phi)}{r^{l+1}}.\label{pr}
\end{eqnarray}
Here $Y_{lm}(\theta, \phi)$ are the spherical harmonics defined by
\begin{eqnarray}&&
Y_{lm}=\sqrt{\frac{(2l+1)}{4\pi}\frac{(l-m)!}{(l+m)!}}P_l^m(\cos\theta)\re^{\ri m \phi}, \nonumber 
\end{eqnarray}
where $P_l^m$ are the associated Legendre polynomials. The term with $l\!=\!0$ is omitted assuming that there is no net mass flux at infinity. The constant coefficients $c_{lm}$ are determined from the boundary conditions at the sphere surface. Assuming arbitrary velocity distribution at the unit sphere we have
\begin{eqnarray}&&
c_{lm}\!=\! \frac{ \lambda\left((l+2) \kappa_l
\!+\!\lambda \kappa_{l-1}\right)}{(l\!+\!1) \kappa_{l-1}}\oint_{r=1}Y_{lm}^* u_r d\Omega
\nonumber\\&&
-\frac{\lambda \kappa_l}{(l\!+\!1) \kappa_{l-1}} \oint_{r=1}  Y_{lm}^* \nabla_s\!\cdot\!\bm u d\Omega,\label{cope}
\end{eqnarray}
where $\kappa_l\!\equiv\!\kappa_l(\lambda^{-1})$ are polynomials in powers of $\lambda^{-1}$ defined via the modified Bessel functions $K_{\nu}(\lambda)$ of the second kind:
\begin{eqnarray}&&\!\!\!\!\!\!\!\!\!\!\!\!
\kappa_l(x)\!\equiv \!\sum_{k=0}^l \frac{(l\!+\!k)!x^k}{k!(l\!-\!k)!2^k},\
K_{l+1/2}(\lambda)\!=\!\sqrt{\!\frac{\pi}{2\lambda}\!}e^{-\lambda}\kappa_l\left(\!\frac{1}{\lambda}\!\right). \label{mod}
\end{eqnarray}
The second integral in (\ref{cope}) contains the surface divergence at $r\!=\!1$ \cite{kim}:
\begin{eqnarray}&&\!\!\!\!\!\!\!
\nabla_s\!\cdot\!\bm u\!=\!\nabla\cdot \bm u\!-\!\frac{\partial u_r}{\partial r}\!=\!2u_r\!+\!\frac{\partial_{\theta}(\sin\theta u_{\theta})\!+\!\partial_{\phi}u_{\phi}}{\sin\theta}, \label{sp}
\end{eqnarray}
where in the first equality $\bm u$ stands for an arbitrary smooth continuation of the surface velocity at $r\!=\!1$ to $r\!>\!1$.

The second term in the solution (\ref{su}) is associated with oscillatory rotations of the boundary at $r\!=\!1$. It satisfies the Helmoholtz equation and the function $X$ can be written as
\begin{eqnarray}&&\!\!\!\!\!\!\!\!\!\!\!\!
X\!=\! \sum_{l=1}^{\infty}\sum_{m=-l}^l {\tilde c}_{lm} \kappa_{l}\left(\frac{1}{\lambda r}\right)Y_{lm}(\theta, \phi), \label{X}
\end{eqnarray}
where the coefficients ${\tilde c}_{lm}$ are obtained from the boundary conditions as
\begin{eqnarray}&&\!\!\!\!\!\!\!\!\!\!\!\!
{\tilde c}_{lm}\!=\!\frac{1}{l(l+1)\kappa_l\left(\lambda^{-1}\right)}\oint_{r=1} Y_{lm}^* (\nabla\!\times\! \bm u)_r   d\Omega. \label{cd}
\end{eqnarray}
The integral in (\ref{cd}) is defined uniquely by the boundary conditions, since the radial component of the curl of the velocity is determined solely by its tangential derivatives.

Finally the remaining $\bm u^H$ term in Eq.~(\ref{su}) is given by
\begin{eqnarray}&&
\bm u^H\!=\!-\!\nabla \re^{\lambda(1-r)}\sum_{l m}\left[\frac{l }{\lambda r}\kappa_{l}\left(\frac{1}{\lambda r}\right)
\!+\!\kappa_{l-1}\left(\frac{1}{\lambda r}\right)\right] {\tilde c}^{r}_{lm} Y_{lm}
\nonumber\\&&
-\re^{\lambda(1-r)}\sum_{l m} \lambda {\tilde c}^{r}_{lm}\bm {\hat r} Y_{lm} \! \left[\frac{4l^2\!-\!1}{2\lambda r} \kappa_{l-1}\!\left(\frac{1}{\lambda r}\right)\!+\!
\kappa_{l-2}\!\left(\frac{1}{\lambda r}\right) \right] ,\nonumber
\end{eqnarray}
where $\sum_{l m}\!\equiv\! \sum_{l=1}^{\infty}\sum_{m=-l}^l$ and
\begin{eqnarray}&&
{\tilde c}^{r}_{lm} \!=\!\frac{\oint_{r=1}  Y_{lm}^* \nabla_s\!\cdot\!\bm u d\Omega-(l+2)\oint_{r=1} Y_{lm}^*u_r d\Omega}{l(l\!+\!1)\kappa_{l-1}(\lambda^{-1})}.  \label{coefficients}
\end{eqnarray}
The resulting net excess shear force is contributed only by the last two terms in Eq.~(\ref{su}), since $\int \partial_x p dxdy=0$, cf. Eqs.~(\ref{fo}) and (\ref{fo1}). By combining the zero and first-order contributions, we conclude that
\begin{eqnarray}&&
F_s\!=\!\re^{-\lambda h}(\partial_h\!-\!\lambda) (C_X\!+\!C^H), \label{ifd}
\end{eqnarray}
where
\begin{eqnarray}
&& C_X\!\equiv\! \int_{z=-h} \left(\nabla\times (\bm {\hat r} \re^{\lambda (1-r)}X )\right)_x dx dy. \nonumber\\
&& C^H\!\equiv \!\int_{z=-h}  u^H_x dxdy. \label{ifq}
\end{eqnarray}
The details of the derivation of coefficients $C_X$ and $C^H$ are provided in Appendices~\ref{app:B} and \ref{app:C}, respectively. The final expressions for these coefficients read:
\begin{eqnarray}
C_X&=&\frac{2 \pi^2 \re^{-\lambda h}}{\lambda^2} \sum_{l=1}^{\infty}\frac{(2l\!+\!1)I_{l+1/2}(\lambda)}{K_{l+1/2}(\lambda) }, \label{cx3} \\
C^H&=&\frac{\pi^2 \re^{-\lambda h}}{\lambda^2} \sum_{l=1}^{\infty}\frac{2(2l\!+\!1)I_{l-1/2}(\lambda)}{K_{l-1/2}(\lambda) }, \label{ch3}
\end{eqnarray}
where $I_{\nu}(\lambda)$ are the modified Bessel functions of the first kind. Substituting $C_X$ from Eq.~(\ref{cx3}) and $C^H$ from Eq.~(\ref{ch3}) into Eq.~(\ref{ifd}) and rearranging the series, we obtain the closed-form expression for the excess shear force exerted on a plate in presence of a distant stationary particle:
\begin{eqnarray}
F_s&=&-\frac{\pi^2 \re^{-2\lambda h} }{\lambda} \left[\frac{3(\re^{2\lambda}-1)}{\pi}
\right.\nonumber\\&& \,\,
\left.+\sum_{l=1}^{\infty}\frac{4(l\!+\!1)I_{l+1/2}(\lambda)}{K_{l+1/2}(\lambda) }\right]\,.
\label{Ffasymp}
\end{eqnarray}
Here we used the identity $I_{1/2}(\lambda)\!=\!\sinh{\lambda\sqrt{2/(\pi\lambda)}}$. The comparison of the approximate solution in Eq.~(\ref{Ffasymp}) to the rigorous numerical results is depicted in Fig.~\ref{fig:Fs}, showing a remarkably close agreement for a wide range of particle sizes $a/\delta$ and arbitrary close proximity to the plate $h/a$.
Expanding the result in Eq.~(\ref{Ffasymp}) for small $a/\delta$ we find
\begin{equation}
F_s=-6\pi\re^{-2 \lambda h} \left[1+\lambda +\frac{14 \lambda^2}{9}+\mathcal{O}(\lambda^3)\right]\,. \label{Ffasymp0}
\end{equation}
This formula provides higher order corrections in the particle size to the expression in Eq.~(\ref{farfs}) obtained in Sec.~\ref{sec:sma} using small-particle limit.

\subsection{Adsorbed and freely suspended particles}

In order to extend the approximate analysis to excess force due to a distant \emph{freely suspended} particle in Eq.~(\ref{Ffree}), one has to derive particle translation and angular velocities $U$ and $\Omega$ and the resistance coefficients $\mathcal A$ and $\mathcal B$, which, in their turn, require components of the resistance matrix $\bm {\mathcal R}$. The excess shear force due to an \emph{adsorbed} particle only requires the knowledge of $\mathcal A$. To the leading approximation in the distance to the plate, the velocity $\bm v$ in Eqs.~(\ref{flo}) is due to a sphere oscillating in the infinite space. Thus, at this order
\begin{eqnarray}&&\!\!\!\!\!\!
{\mathcal R}_{11}=-6\pi\left(1+\lambda+\frac{\lambda^2}{9}\right),\ \ {\mathcal R}_{12}=0, \label{R11asymp}
\end{eqnarray}
where ${\mathcal R}_{11}$ and ${\mathcal R}_{12}$ are the corresponding force and torque exerted on the sphere undergoing oscillatory translations with velocity $\bm \hat{x}$  in unbounded viscous fluid \cite{kim}
[for higher order corrections to ${\mathcal R}_{11}$ see Eq.~(\ref{R11})].
The calculation of $\mathcal A$ follows the similar approach to that in the calculation of $F_s$. We write $\bm v=\bm v^0+\bm v^1$, where $\bm v^0$ is the well-known flow due to sphere oscillating in the infinite space \cite{LL,kim} that can be written as \cite{fouxon22}
\begin{eqnarray}&&\!\!\!\!\!\!
\bm v^0=\bm v_s-\frac{\nabla p}{\lambda^2}, \ \ p=-\frac{3}{2}\left(1\!+\!\lambda\!+\!\frac{\lambda^2}{3}\right)  (\bm {\hat x}\cdot \nabla)\frac{1}{r},
\end{eqnarray}
where $\bm v_s$ is the solution of the vector Helmholtz equation given by
\begin{eqnarray}&&\!\!\!\!\!\!
\bm v_s=\left(\bm {\hat x}\nabla^2 - \partial_x \nabla \right) \frac{3 \re^{\lambda(1-r)}}{2\lambda^2 r}.
\end{eqnarray}
The calculation analogous to that leading to Eqs.~(\ref{fo}) and (\ref{fo1}), shows that $\mathcal A$ defined in Eqs.~(\ref{flo})-(\ref{s}) is given by
\begin{eqnarray}
&&\mathcal A=\left(\lambda-\partial_h\right)\int_{z=-h} v^0_xdxdy=(\lambda-\partial_h)\nonumber\\
&& \times \int_{z=-h} \nabla^2\left[\frac{3\re^{\lambda(1-r)}}{2\lambda^2 r}\right] dxdy=
\frac{3\re^\lambda}{2\lambda^2} (\lambda-\partial_h)\partial_h^2 {\mathcal J},\, \label{i1}
\end{eqnarray}
where we introduced
\begin{eqnarray}&&
\mathcal{J}\!\equiv \!\int_{z=-h}\! \frac{\re^{-\lambda r}}{r} dxdy
\!=\!2\pi h\! \int_{\pi/2}^{\pi}\! \frac{\tan{\theta}}{\cos{\theta}}\, \re^{\lambda h/\cos{\theta}} d\theta ,\nonumber
\end{eqnarray}
and applied the transformation in Eq.~(\ref{transofm}). Using the calculation similar to that in the previous sections it follows that
$\mathcal {J}\!=\!2\pi \re^{-\lambda h}/\lambda$. Substituting the last result into (\ref{i1}) readily gives
\begin{eqnarray}&&
\mathcal A=6\pi \re^{\lambda(1-h)}. \label{Aasymp}
\end{eqnarray}
The approximate expression for the excess shear force due to an adsorbed particle can readily be obtained from Eq.~(\ref{Fa}):
\begin{eqnarray}&&
F_a=F_s+6\pi\, \re^{\lambda(1-h)}\,, \label{Faasymp}
\end{eqnarray}
where $F_s$ is given by Eq.~(\ref{Ffasymp}). Expanding $\mathcal A$ in (\ref{Aasymp}) for small $a/\delta$ and using (\ref{Ffasymp0}) we obtain
\begin{eqnarray}
F_a&=&-6\pi \re^{-2 \lambda h} \left(1+\lambda +\frac{14\lambda^2}{6}\right) \nonumber \\
&& +6\pi \re^{-\lambda h} \left(1+\lambda +\frac{\lambda^2}{2}\right)+ \mathcal{O}(|\lambda|^3)\,. \label{Faasymp0}
\end{eqnarray}
In the leading order in $\lambda$ the Eq.~(\ref{Faasymp0}) is identical to the expression (\ref{Fafar}) obtained in Sec.~\ref{sec:sma} using small-particle approximate theory.

We next determine the asymptotic expressions for ${\mathcal R}_{22}$ and $\mathcal B$. To the leading approximation in separation distance, ${\mathcal R}_{22}$ is equal to the torque exerted on the sphere undergoing oscillatory rotations in unbounded viscous fluid with angular velocity $\bm {\hat y}$ \cite{kim}:
\begin{eqnarray}&&
\mathcal{R}_{22}=-8\pi-\frac{8\pi \lambda^2}{3(1+\lambda)}. \label{R22asymp}
\end{eqnarray}
The corresponding flow velocity reads:
\begin{eqnarray}&&
\bm v^{0r}\!=\!\frac{(1\!+\!\lambda r)\bm {\hat y}\times\bm r}{(1\!+\!\lambda)r^3} \re^{-\lambda(r-1)}.
\end{eqnarray}
We further find that $\mathcal B$ defined in Eq.~(\ref{t}) obeys
\begin{eqnarray}&&
{\mathcal B}=\left(\lambda-\partial_h\right)\int_{z=-h} v^{0r}_x dxdy=
\nonumber\\&&
\left(\lambda-\partial_h\right)\int_{z=-h} \frac{(1\!+\!\lambda r)z}{(1\!+\!\lambda)r^3}\, \re^{-\lambda(r-1)} dxdy.
\end{eqnarray}
Applying the transformation in Eq.~(\ref{transofm}) we readily find
\begin{eqnarray}&&
{\mathcal B}\approx \left(\lambda-\partial_h\right) \frac{2 \pi h \lambda\re^\lambda}{1\!+\!\lambda}
\int_{\pi/2}^{\pi} \tan{\theta}\, \re^{\lambda h/\!\cos{\theta}}\, d\theta  \nonumber\\&&
= \left(\partial_h-\lambda\right) \frac{2 \pi \re^{\lambda(1-h)}}{1+\lambda}=-\frac{4 \pi \lambda \re^{\lambda(1-h)}}{1+\lambda}\,. \label{Basymp}
\end{eqnarray}
Substituting the derived asymptotic expressions for $\mathcal A$, $\mathcal B$ in Eqs.~(\ref{Aasymp}) and (\ref{Basymp}), and the components of the resistance matrix in Eqs.~(\ref{R11asymp}) and (\ref{R22asymp}) into Eqs.~(\ref{motion}) results in compact closed-form expressions for the approximate translation and rotation velocities of the freely-suspended particle of arbitrary mass located at distance $h$ above the oscillating plate:
\begin{eqnarray}&&
V \approx \frac{18\pi\,\re^{\lambda(1-h)}}{2\pi(9\!+\!9\lambda+\!\lambda^2)+3\lambda^2\xi},
\label{Vasymp0}\\
&&
\Omega \approx -\frac{30\pi\lambda\, \re^{\lambda(1-h)}}{20\pi(3\!+\!3\lambda+\!\lambda^2)+3\lambda^2(1\!+\!\lambda)\xi}. \label{Omasymp0}
\end{eqnarray}
Expanding the expression $\mathcal{A}V+\mathcal{B}\Omega$ showing in the Eq.~(\ref{Fdef}) for the excess shear force due to a free suspended particle for small $a/\delta$ up to the second order in $\lambda$, we find that
\be
\mathcal{A}V+\mathcal{B}\Omega\!=\!\re^{-2\lambda h}\left[6\pi+6\pi\lambda+\frac{1}{3}\lambda^2 (22\pi-3\xi)+\mathcal{O}(|\lambda|^3) \right]\,.\nonumber
\ee
Combining with result with the analogous expansion of $F_s$ in Eq.~(\ref{Ffasymp0}) we find the expression for excess shear force due to small ($a/\delta \ll 1$) freely suspended particle:
\be
F\approx -\lambda^2 \re^{-2\lambda h}\left(2\pi+\xi \right)\,. \label{Fasymp0}
\ee
Notice that while the distant- and the small-particle theory of Sec.~\ref{sec:sma} yield the same result for $F_a$
and $F_s$ in the limit $a\!\ll\!\delta$ to the leading approximation, the distant-particle theory also provides the leading-order result for $F$ which requires higher-order expansion in $\lambda$, which was not readily available from the small-particle theory.

For a neutrally buoyant particle ($\xi\!=\!4\pi/3$) the Eqs.~(\ref{Vasymp0})--(\ref{Omasymp0}) become:
\begin{eqnarray}&&
V \approx \frac{3\,\re^{\lambda(1-h)}}{3\!+\!3\lambda+\!\lambda^2},
\label{Vasymp}\\
&&
\Omega \approx -\frac{15\lambda\, \re^{\lambda(1-h)}}{10(3\!+\!3\lambda+\!\lambda^2)+2(1\!+\!\lambda)\lambda^2}. \label{Omasymp}
\end{eqnarray}
Notice that in the limit of small particle, $a/\delta \ll 1$, we readily obtain that to the first approximation it translates with the velocity of the undisturbed flow, $V\approx \re^{-\lambda h}$ and rotates with angular velocity $\Omega\approx \lambda \re^{-\lambda h}/2$. The result for $V$ reproduces Eq.~(\ref{farvl}), as opposed to the leading-order expression for $\Omega$ which was unavailable in the framework of small-particle theory in Sec.~\ref{sec:sma}.

Substituting $V$ and $\Omega$ in Eqs.~(\ref{Vasymp})--(\ref{Omasymp}) together with (\ref{Aasymp}) and (\ref{Basymp}) into (\ref{Fdef}) produces the asymptotic expression for the excess shear force due to a freely suspended neutrally buoyant distant particle:
\begin{eqnarray}
F&=&F_s+ 6\pi \re^{2 (1-h) \lambda} \left[\frac{3}{3+\lambda (3+\lambda)} \right. + \nonumber \\
&& \left.\frac{5\lambda^2}{(1+\lambda) (15+\lambda (15+\lambda (6+\lambda)))}
\right]\,, \label{Fasymp}
\end{eqnarray}
where $F_s$ is given by Eq.~(\ref{Ffasymp}). In the small-particle limit (for $a/\delta \!\ll\! 1$) we thus have $F\approx -10\pi \lambda^2\re^{-2\lambda h}/3 $, in agreement with (\ref{Fasymp0}).

As for a stationary particle, the comparison of the derived approximate solutions for firmly adhered and freely suspended particles in Eqs.~(\ref{Faasymp}) and (\ref{Fasymp}) to the results of FEM computations is depicted in Figs.~\ref{fig:Fa} and \ref{fig:Ffree}, respectively, showing a very close agreement for a wide range of particle sizes $a/\delta$ and separation distances $h/a$. Comparison with the numerical results also demonstrates that the simple formula (\ref{Fasymp0}) provides an accurate prediction for the excess shear force due to a small particle with radius $a\! \lesssim\! 0.25\delta$ located at $h \gtrsim a$ from the oscillating plate (see Sec.~\ref{sec:res}).

\section{Close proximity and lubrication theory \label{sec:lubr}}
In this Section we consider the asymptotic limit of vanishing separation between the particle and the oscillating plate. This limit is of a practical interest to the QCM-D devices designed to measure impedance due to particles adsorbed at the resonator surface. To understand the role of the hydrodynamic forces to the excess shear force exterted on the plate, we follow Ref.~\cite{Busca21} and entirely neglect the contribution of nonhydrodynamic (adhesion) forces. However, as opposed to Ref.~\cite{Busca21}, we do not make any \emph{a priori} assumptions concerning the fluid-mediated motion of the freely suspended particle and determine the velocity of the particle self-consistently from the solution.

In the limit of zero oscillation frequency, the steady Stokes equations apply and the limit of vanishing separations is covered by the classical lubrication theory. The minimal gap between the particle surface and the plane is given by $h-a\!=\!\epsilon a$, where $\epsilon \ll 1$. The lubrication theory seeks the solution as asymptotic series expansion in the small parameter $\epsilon$ (see, e.g., \cite{kim}). Here we would like
extend the classical lubrication theory to transient Stokes equations.

Let us consider the case of a freely suspended particle. The aim is to determine the velocities provided by Eq.~(\ref{motion}) in the limit of $\epsilon\!\to\!0$. This requires derivation of the resistance coefficients ${\mathcal R}_{ik}$, $\mathcal A$ and $\mathcal B$ in Eq.~(\ref{motion}). These coefficients can be obtained by analyzing the flow given by Eqs.~(\ref{flo}) and the same flow with the boundary condition $\bm v(r\!=\!1)=\hat{\bm x}$ replaced by $\bm v(r\!=\!1)=\hat{\bm y}\times \bm r$, see remarks after Eqs.~(\ref{flo}). Both cases can be treated similarly and in cylindrical coordinates $(\varrho,\phi,z)$ the pressure and the velocity components satisfy the following equations \cite{LL}:
\begin{eqnarray}&&\!\!\!\!\!\!\!\!\!\!\!\!\!
\lambda^2 v_{\varrho}\!=\!-\frac{\partial p}{\partial \varrho}\!+\!\nabla^2 v_{\varrho}-\frac{v_{\varrho}}{\varrho^2}-\frac{2}{\varrho^2} \frac{\partial v_\phi}{\partial \phi},\nonumber\\&& \!\!\!\!\!\!\!\!\!\!\!\!\!
\lambda^2 v_{\phi}\!=\!-\frac{1}{\varrho}\frac{\partial p}{\partial \phi}\!+\!\nabla^2 v_{\phi}-\frac{v_{\phi}}{\varrho^2}+\frac{2}{\varrho^2}\frac{\partial v_{\varrho}}{\partial\phi},\nonumber\\&& \!\!\!\!\!\!\!\!\!\!\!\!\!
\lambda^2 v_z\!=\!-\frac{\partial p}{\partial z}\!+\!\nabla^2 v_z, \label{cylind}
\end{eqnarray}
where
\begin{eqnarray}&&\!\!\!\!\!\!\!\!\!\!\!\!\!
\nabla^2=\frac{1}{\varrho}\frac{\partial}{\partial \varrho} \left(\varrho\frac{\partial}{\partial \varrho}\right)+\frac{1}{\varrho^2}\frac{\partial^2}{\partial \phi^2}+\frac{\partial^2}{\partial z^2},
\end{eqnarray}

Further progress can be made by noting the dependence of the solution on the polar angle $\phi$ \cite{fl18}:
\begin{eqnarray}&&\!\!\!\!\!\!\!\!\!\!\!\!\!
v_{\varrho}\!=\!U(\varrho, z)\cos\phi, \ \ v_{\phi}\!=\!V(\varrho, z)\sin\phi,\nonumber\\&& \!\!\!\!\!\!\!\!\!\!\!\!\!
v_z\!=\! W(\varrho, z)\cos\phi,\ \
p={\tilde P}(\varrho, z)\cos\phi\,, \label{as}
\end{eqnarray}
where $U$, $V$, $W$ and $P$ are functions of the radial, $\varrho$, and axial, $z$, cylindrical coordinates. Substituting the ansatz (\ref{as}) into the Eqs.~(\ref{cylind}) yields:

\begin{eqnarray}&&\!\!\!\!\!\!\!\!\!\!\!\!\!
\lambda^2 U\!=\!-\partial_{\varrho} {\tilde P}\!+\!L_0^2 U-\frac{2(U+V)}{\varrho^2},\nonumber\\&& \!\!\!\!\!\!\!\!\!\!\!\!\!
\lambda^2  V\!=\!\frac{{\tilde P}}{\varrho}\!+\!L_0^2 V-\frac{2(U+V)}{\varrho^2},\nonumber\\&& \!\!\!\!\!\!\!\!\!\!\!\!\!
\lambda^2  W\!=\!-\partial_z {\tilde P}\!+\!L_1^2W, \label{lubr}
\end{eqnarray}
where we defined the operator $L_m^2$ as
\begin{eqnarray}&&\!\!\!\!\!\!\!\!\!\!\!\!\!
L_m^2=\frac{\partial^2}{\partial \varrho^2}+\frac{1}{\varrho}\frac{\partial}{\partial \varrho}-\frac{m^2}{\varrho^2}+\frac{\partial^2}{\partial z^2}\,.
\end{eqnarray}
The continuity equation gives
\begin{eqnarray}&&\!\!\!\!\!\!\!\!\!\!\!\!\!
\frac{\partial U}{\partial \varrho}+\frac{V+U}{\varrho}+\frac{\partial W}{\partial z}=0. \label{lubr_cont}
\end{eqnarray}
To implement small-$\epsilon$ expansion we proceed as along the same lines of the standard lubrication theory \cite{kim}. We re-scale the coordinates $(\varrho, z)$ according to
\begin{eqnarray}&&\!\!\!\!\!\!\!\!\!\!\!\!\!
R=\frac{\varrho}{\sqrt{\epsilon}},\ \ Z=\frac{z}{\epsilon}.
\end{eqnarray}
In the re-scaled coordinates the Eqs.~(\ref{lubr})--(\ref{lubr_cont}) then become
\begin{eqnarray}
&&\!\! \lambda^2 U\!=\!-\frac{1}{\sqrt{\epsilon}}\frac{\partial P}{\partial R}\!-\!\frac{2(U+V)}{R^2\epsilon}
\!+\!\frac{1}{\epsilon}\frac{\partial^2 U}{\partial R^2}\!+\!\frac{1}{R\epsilon}\frac{\partial U}{\partial R}\!+\!\frac{1}{\epsilon^2}\frac{\partial^2 U}{\partial Z^2}
,\nonumber  \\ 
&& \!\! \lambda^2 V\!=\!\frac{P}{\sqrt{\epsilon} R}\!-\!\frac{2(U+V)}{R^2 \epsilon}
+\frac{1}{\epsilon}\frac{\partial^2 V}{\partial R^2}\!+\!\frac{1}{R\epsilon}\frac{\partial V}{\partial R}\!+\!\frac{1}{\epsilon^2}\frac{\partial^2 V}{\partial Z^2},\nonumber \\ 
&& \!\! \lambda^2 W\!=\!-\!\frac{1}{\epsilon}\frac{\partial P}{\partial Z}\!-\!\frac{W}{R^2 \epsilon}
\!+\!\frac{1}{\epsilon}\frac{\partial^2 W}{\partial R^2}\!+\!\frac{1}{R\epsilon}\frac{\partial W}{\partial R}\!+\!\frac{1}{\epsilon^2}\frac{\partial^2 W}{\partial Z^2},
\nonumber \\ 
&& \frac{\partial U}{\partial R}\!+\!\frac{V+U}{R}\!+\!\frac{1}{\sqrt{\epsilon}}\frac{\partial W}{\partial Z}=0. \label{saop}
\end{eqnarray}
Notice that the coefficient of the transient term does not contain $\epsilon$. This suggests that unsteadiness does not affect the solution either at the leading (zero) order or the sub-leading (first) order in $\epsilon$, provided that $|\lambda|\epsilon \ll 1$. In other words, for small clearances $\epsilon$ satisfying $\epsilon \!\ll\! \delta/a$, the two-term asymptotic expansion of the solution has the same form as the solution of the steady Stokes equations \cite{kim}:
\begin{eqnarray}&&\!\!\!\!\!\!\!\!\!\!\!\!\!
P(R, Z)=\epsilon^{-3/2}P_0(R, Z)+\epsilon^{-1/2}P_1(R, Z)+\ldots,\nonumber\\&&\!\!\!\!\!\!\!\!\!\!\!\!\!
U(R, Z)=U_0(R, Z)+\epsilon U_1(R, Z)+\ldots,\nonumber\\&&\!\!\!\!\!\!\!\!\!\!\!\!\!
V(R, Z)=V_0(R, Z)+\epsilon V_1(R, Z)+\ldots,\nonumber\\&&\!\!\!\!\!\!\!\!\!\!\!\!\!
W(R, Z)=\epsilon^{1/2}W_0(R, Z)+\epsilon^{3/2}W_1(R, Z)+\ldots,
\end{eqnarray}
Thus at the leading order the flow is governed by the equations :
\begin{eqnarray}&&\!\!\!\!\!\!\!\!\!\!\!\!\!
\frac{\partial P_0}{\partial R}=\frac{\partial^2 U_0}{\partial Z^2},\ \ -\frac{P_0}{R}=\frac{\partial^2 V_0}{\partial Z^2},\ \
\frac{\partial P_0}{\partial Z}=0,\nonumber\\&&\!\!\!\!\!\!\!\!\!\!\!\!\!
\frac{\partial U_0}{\partial R}+\frac{V_0+U_0}{R}+\frac{\partial W_0}{\partial Z}=0. \label{lowestorder}
\end{eqnarray}
The sub-leading (first) order solution satisfies:
\begin{eqnarray}&&\!\!\!\!\!\!\!\!\!\!\!\!\!
\frac{\partial P_1}{\partial R}\!=\!\frac{\partial^2 U_1}{\partial Z^2}-\frac{2(U_0+V_0)}{R^2}
+\frac{\partial^2 U_0}{\partial R^2}+\frac{1}{R}\frac{\partial U_0}{\partial R},\nonumber\\&& \!\!\!\!\!\!\!\!\!\!\!\!\!
-\frac{P_1}{R}\!=\!\frac{\partial^2 V_1}{\partial Z^2}-\frac{2(U_0+V_0)}{R^2}
+\frac{\partial^2 V_0}{\partial R^2}+\frac{1}{R}\frac{\partial V_0}{\partial R}
,\nonumber\\&& \!\!\!\!\!\!\!\!\!\!\!\!\!
\frac{\partial P_1}{\partial Z}\!=\!\frac{\partial^2 W_0}{\partial Z^2};\ \
\frac{\partial U_1}{\partial R}+\frac{V_1+U_1}{R}+\frac{\partial W_1}{\partial Z}=0. \label{lowestorder1}
\end{eqnarray}
These are the same lubrication equations that hold for the steady Stokes equations (see Ch.~9 in Ref.~\cite{kim}). Apparently, the unsteadiness affects the solution only at the next (second) order in $\epsilon$.

We conclude that we can apply the results of the standard lubrication theory of the usual Stokes equations in the first two orders. We find using the results for the force and the torque due to shearing motions of rigid surfaces, see, e.g.,  \cite{kim}, that the $2\times 2$ resistance matrix $\bm {\mathcal R}$ with elements ${\mathcal R}_{ik}$ is given by
\begin{eqnarray}&&\!\!\!\!\!\!\!\!\!\!\!\!\!
\bm {\mathcal R}=-\frac{4\pi}{5}\ln{\epsilon^{-1}}\,
\left( \begin{array}{cc}
4  & -1 \\
-1 & 4
\end{array} \right), \label{amhaz}
\end{eqnarray}
where we used representations of ${\mathcal R}_{ik}$ via forces and torques given by Eqs.~(\ref{s})-(\ref{t}). The next order corrections to forces and torques are of $\mathcal{O}(1)$ \cite{kim}. Hence the above two-term expansion is accurate provided that separation distance is exponentially small, $|\ln\epsilon|\gg 1$. Recall that there is an additional condition $\epsilon \ll \delta/a$ that guarantees that the
unsteadiness is not affecting the two-term expansion of the solution.

In order to obtain the coefficients $\mathcal A$ and $\mathcal B$ in Eq.~(\ref{motion}) we use their representations in Eqs.~(\ref{os}) and (\ref{Bdefsph}), respectively. We focus on the integral terms in these representations since contribute to the dominant (singular) component of $\mathcal A$ and $\mathcal B$ at vanishing separation, $\epsilon\to 0$. We first consider $\oint _{r=1}\!\!\sigma^u_{xr}dS$ in Eq.~(\ref{os}) in the limit $a\!\ll \!\delta$. In this limit we can approximate $\re^{-\lambda z}$ in the boundary conditions in Eqs.~(\ref{fon17}) by the constant factor $\re^{-\lambda h}$, which reduces the fluid velocity $\bm u$ in Eqs.~(\ref{fon17}) to $-\re^{-\lambda h}\bm v$, where $\bm v$ satisfies Eqs.~(\ref{flo}). By comparing with the definition of ${\mathcal R}_{11}$, we find that ${\mathcal A}=-\re^{-\lambda h}{\mathcal R}_{11}$ in the limit of $\epsilon\to 0$. Similarly we have ${\mathcal B}=-\re^{-\lambda h}{\mathcal R}_{12}$. Then it follows from Eq.~(\ref{motion}) that at $\epsilon\to 0$ we have
\begin{eqnarray}&&\!\!\!\!\!\!\!\!\!\!\!\!\!
\left(\begin{array}{cc}
 V \\
\Omega  \end{array}\right)=\re^{-\lambda h}\frac{1}{{\mathcal R}_{11}{\mathcal R}_{22}-{\mathcal R}_{12}^2}
\nonumber\\&&\!\!\!\!\!\!\!\!\!\!\!\!\!
\times \left(\begin{array}{cc}
{\mathcal R}_{22}   & -{\mathcal R}_{12} \\
-{\mathcal R}_{12} &  {\mathcal R}_{11}\end{array}\right)\left(\begin{array}{cc}
{\mathcal R}_{11} \\
{\mathcal R}_{12} \end{array}\right)\rightarrow \left(\begin{array}{cc}
1\\
0 \end{array}\right),
\end{eqnarray}
where in the limit $\epsilon\!\to\! 0$ we have $h\!\to\! 1$ and $\re^{-\lambda h}\!\to\!1$. Notice that the above derivation holds provided that $|\ln\epsilon|\gg \max{(\mathrm{St}, 1)}$ which may limit its applicability to extremely small $\epsilon$, where non-hydrodynamic (e.g., adhesion) forces are dominant.

While the above derivation also required $a\ll \delta$, the analysis probably holds for an arbitrary value of $a/\delta$, as the singular component of the solution is determined by the velocity of particle surface closest to the plane, such that $\re^{-\lambda z}$ can be replaced by $\re^{-\lambda\epsilon}$. Thus one can conclude that at vanishing separations, in general, the particle velocities satisfy $V\!\to\! 1$ and $\Omega\!\to\! 0$. This argument was used in Ref.~\cite{Busca21} to justify the assumption $F\!\simeq \!F_a$ in the limit $\epsilon\!\to\! 0$.
This assumption is however inaccurate, as shown below. Using Eqs.~(\ref{Fdef}) and (\ref{Fa}) the excess shear force due to a freely suspended particle can be written as
\[
F\!=\!F_a+(V-1){\mathcal A}+\Omega  {\mathcal B}\,,
\]
where $F_a$ corresponds to an adsorbed particle (i.e., oscillating with a plane as a whole) and is therefore finite at constant, $\epsilon\!=\!0$. Although $V\!\to\! 1$, $\Omega\!\to\! 0$, the contributions of the corresponding terms to $F$ at $\epsilon \to 0$ are finite, since both ${\mathcal A}$ and ${\mathcal B}$ diverge at contact. At the leading order in $\epsilon$ we have
\begin{eqnarray}&&
V-1\approx \frac{C_1}{|\ln\epsilon|},\ \ \Omega \approx \frac{C_2}{|\ln\epsilon|},
\end{eqnarray}
where $C_i$ are some constants. By using ${\mathcal A}\approx-\re^{-\lambda h}{\mathcal R}_{11}$ and ${\mathcal B}\approx-\re^{-\lambda h}{\mathcal R}_{12}$ together with Eqs.~(\ref{amhaz}) we readily find that
\begin{eqnarray}&&
\lim_{\epsilon\to 0}F=\left.F_a\right|_{\epsilon=0}-\frac{4 \pi}{5} (4C_1-C_2) \re^{-\lambda}. \label{FvsFa}
\end{eqnarray}
Therefore, unless $|\lambda| \gg 1$, the assumption $F\simeq F_a$ is inaccurate although the ``no-slip" conditions (i.e., $V\to 1$, $\Omega\to 0$) hold to the leading approximation. Finding the constants $C_i$ would typically require matching of the lubrication approximation with accurate numerical solution in the same way as it is done in standard lubrication theory of steady Stokes flows (see, e.g., \cite{kim}). This, however, is beyond the scope of the present paper and will be conducted elsewhere.

\section{Numerical computations \label{sec:num}}
The numerical solution of Eqs.~(\ref{fon17}) is performed in the cylindrical coordinates $\{\varrho,\phi,z\}$
such that $x=\varrho \cos\phi$, $y=\varrho \sin\phi$, $z=z$, with its
origin at the plate and the $z$-axis coinciding with the vertical axis of the spherical particle.
\begin{figure*}[tbh]
\begin{tabular}{cc}
\includegraphics[width=0.35\textwidth]{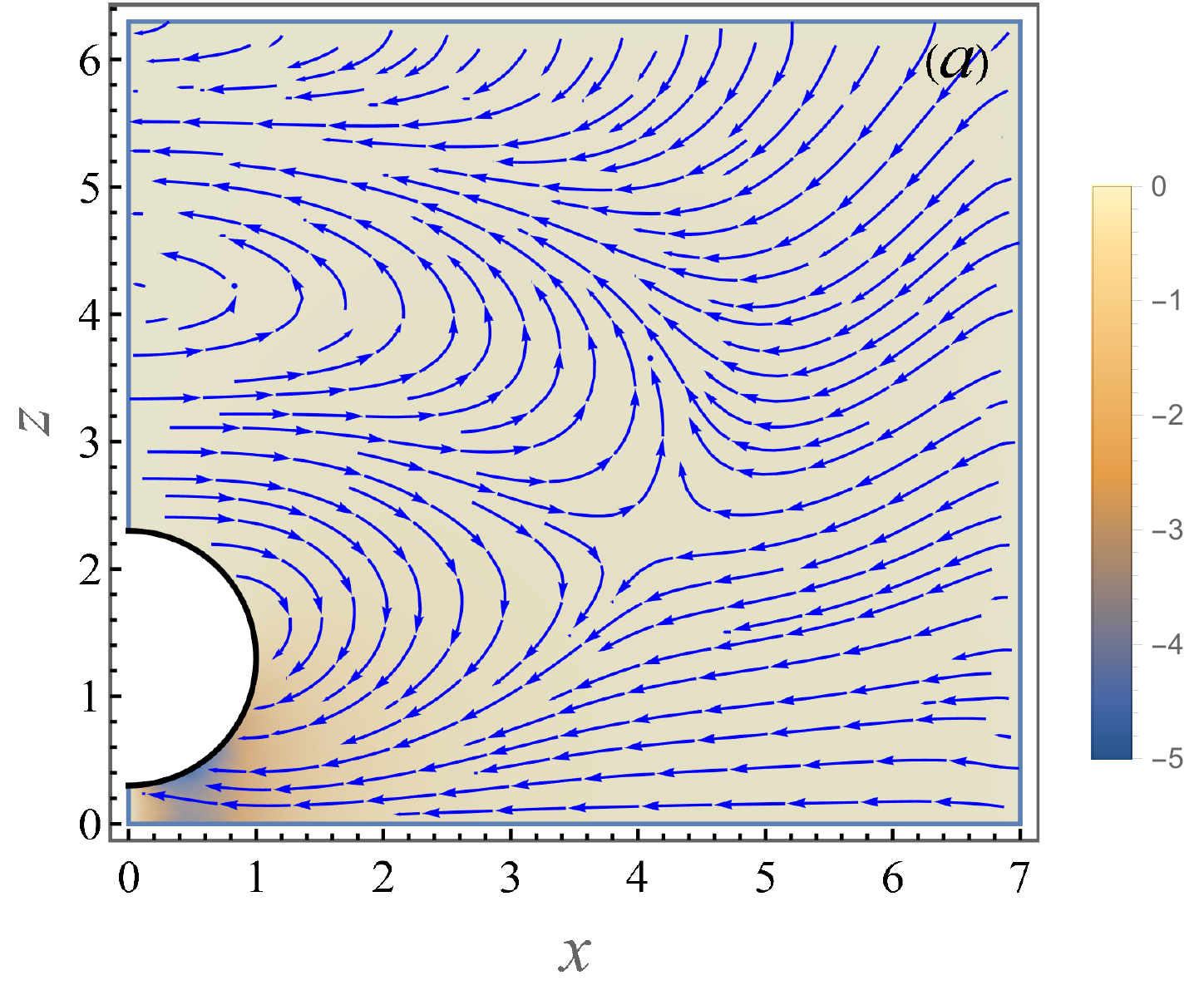}  &
\includegraphics[width=0.355\textwidth]{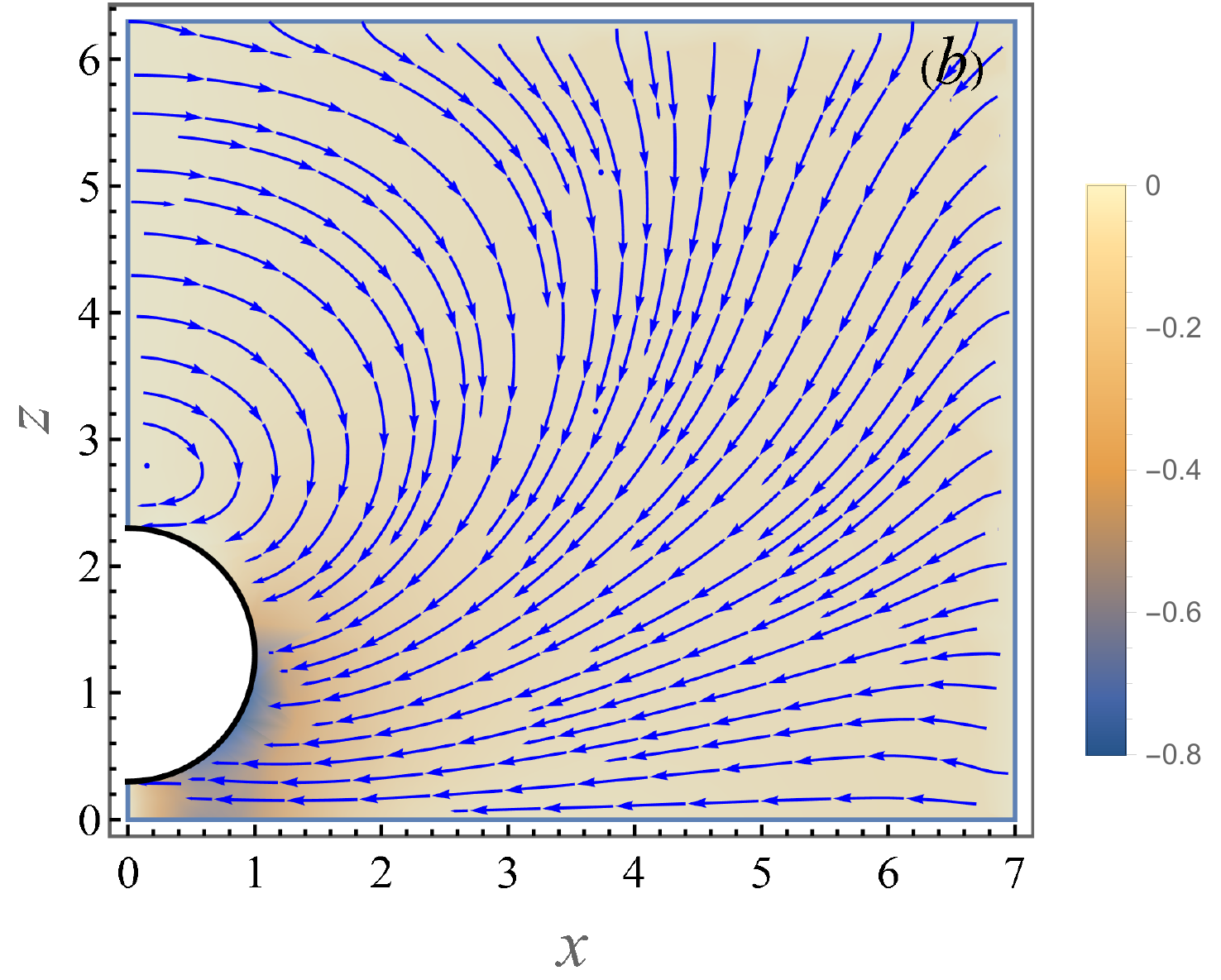} \\ \\
\includegraphics[width=0.355\textwidth]{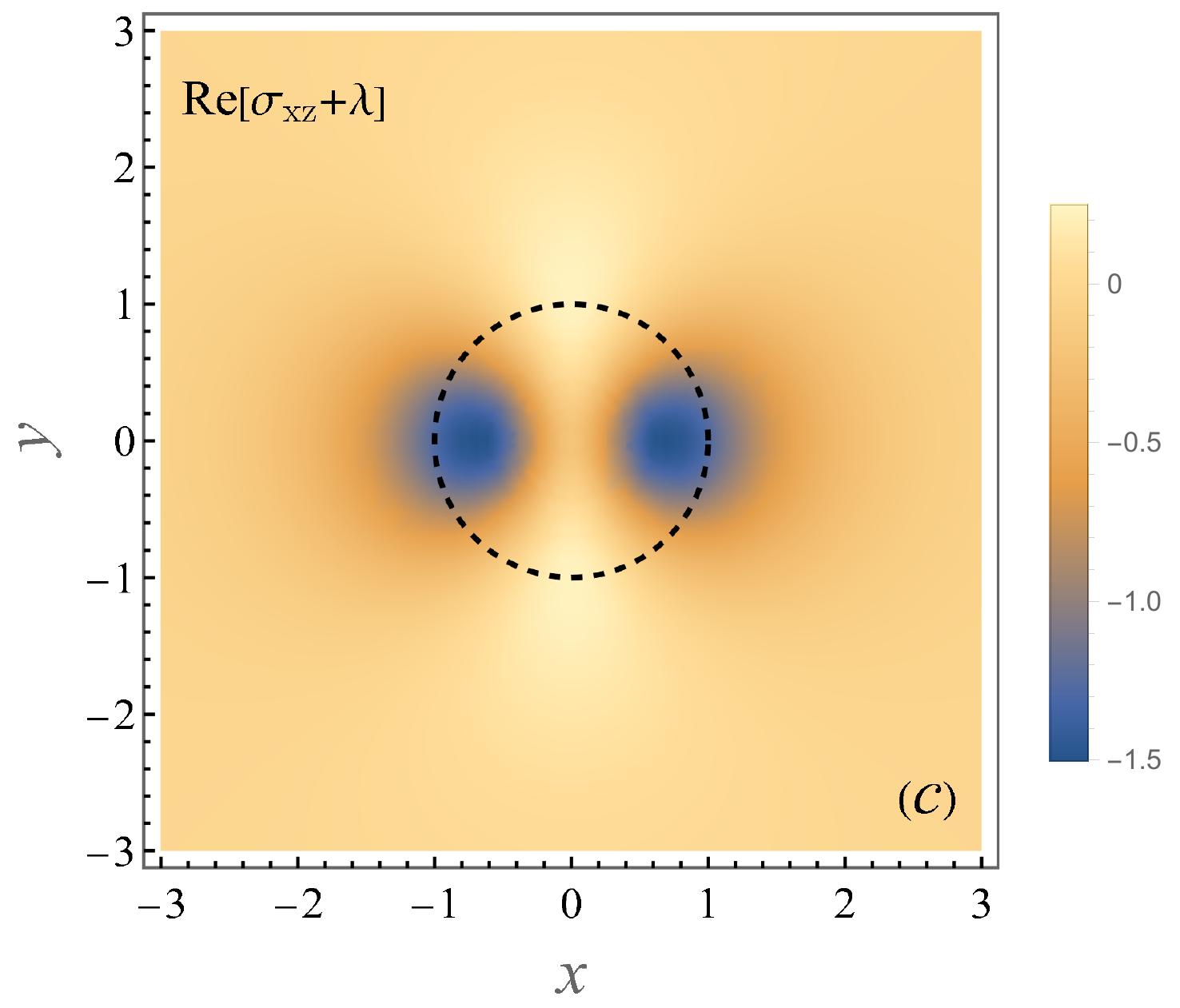}  &
\includegraphics[width=0.355\textwidth]{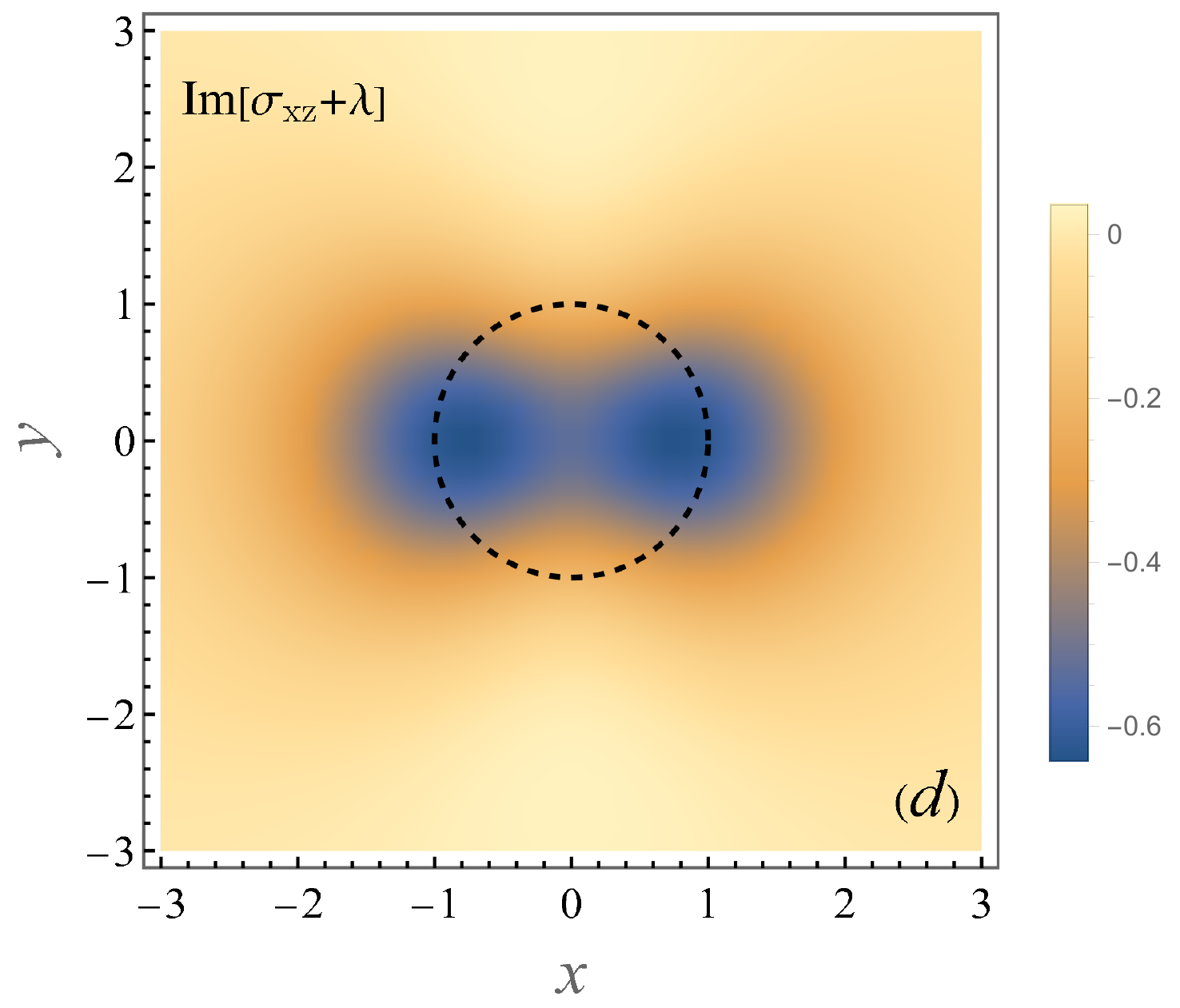}
\end{tabular}
\caption{The upper panel shows the flow and pressure (color map) perturbation fields in Eq.~(\ref{fon17}) due to a stationary particle for $a/\delta\!=\!1$ and $h/a\!=\!1.3$: (a) velocity field $\{\rRe [U], \rRe [W]\}$ and pressure $\rRe[\tilde P]$ at $\omega t=0$; (b) velocity field $\{\rIm [U], \rIm [W]\}$ and pressure $\rIm[\tilde P]$ at $\omega t=\pi/2$. The lower panel shows the corresponding perturbation to the shear stress exerted at the plate at $z\!=\!0$ (color map): c) $\rRe[\sigma_{xz}+\lambda]$ at $\omega t=0$; d) $\rIm[\sigma_{xz}+\lambda]$ at $\omega t=\pi/2$. The dashed circles mark the position of the particle above the plane.  \label{fig:num}}
\end{figure*}
We use the ansatz (\ref{as}) which allows to reduce the solution to two dimensional functions $U$, $V$, $W$ and $\tilde P$ defined in the plane $(\varrho,z)$ bounded by $z\!=\!0$, $z\!=\!z_\mathrm{max}(h)\!>\!0$, $\varrho\!=\!0$, $\varrho\!=\!\varrho_\mathrm{max}>0$ and $\varrho^2+(z-h)^2\!=\!1$.
The pressure $\tilde P$ is set to a fixed (zero) value far from the particle at $z\!=\!z_\mathrm{max},\ \varrho\!=\!\varrho_\mathrm{max}$.
The boundary condition $\bm u\!=\!0$ is used at $\varrho\!=\!\varrho_{max}$,  $z\!=\!0$ and $z\!=\!z_\mathrm{max}$.
We set no-flux boundary condition at $\varrho\! =\!0$, while at the half-circle representing the particle surface. the boundary condition depends on type of the problem (i.e., stationary, firmly adhered or freely suspended particle). In particular, for the computation of $F_s$ using (\ref{forcef}) we specify $U\! =\! - V\!=\! -\re^{-\lambda z}$ and $W \!=\! 0$.
For the computation of $\mathcal A$ and ${\mathcal R}_{11}$, ${\mathcal R}_{12}$ with (\ref{os}) and (\ref{s}) we set
$ U \!=\! - V=1$, $W\! =\! 0$, while finding $\mathcal B$ and ${\mathcal R}_{22}$ with (\ref{Bdefsph}) and (\ref{t})
requires $U\! =\! - V\!=\!z-h$ and $W\! =\! -\varrho$. We use the Finite Element Method (FEM) implemented in {\it Mathematica} 12.0.
A typical mesh size is selected to be $0.05$ within the domain and $0.025$ along its boundaries.

Numerical simulations show that the flow $\bm u$ converges at $\varrho_\mathrm{max} \sim 5-10, z_\mathrm{max} \sim (6-8)+h$. The typical flow and pressure disturbance due to a stationary particle for $a/\delta\!=\!1,$ $h\!=\!1.3$ in $xz$--plane (for $\phi\!=\!0$) are shown in Figs.~\ref{fig:num}a,b at two instances, $\omega t\!=\!0$ and $\omega t\!=\!\pi/2$, respectively. The corresponding distributions of the shear stress perturbation at the $xy$--plane, $\sigma_{xz}+\lambda$, due to presence of the stationary particle are depicted in Figs.~\ref{fig:num}c,d. It can be readily seen, that the interaction of the transverse wave originated at the oscillating plate (see the undisturbed velocity in Fig.~\ref{fig:schematic}) with the particle located above it creates a rather complex flow pattern with ``breathing" recirculations.

\section{Numerical results and comparison to theory \label{sec:res}}
\begin{figure*}[tbh]
\begin{center}
\includegraphics[width=0.70\textwidth]{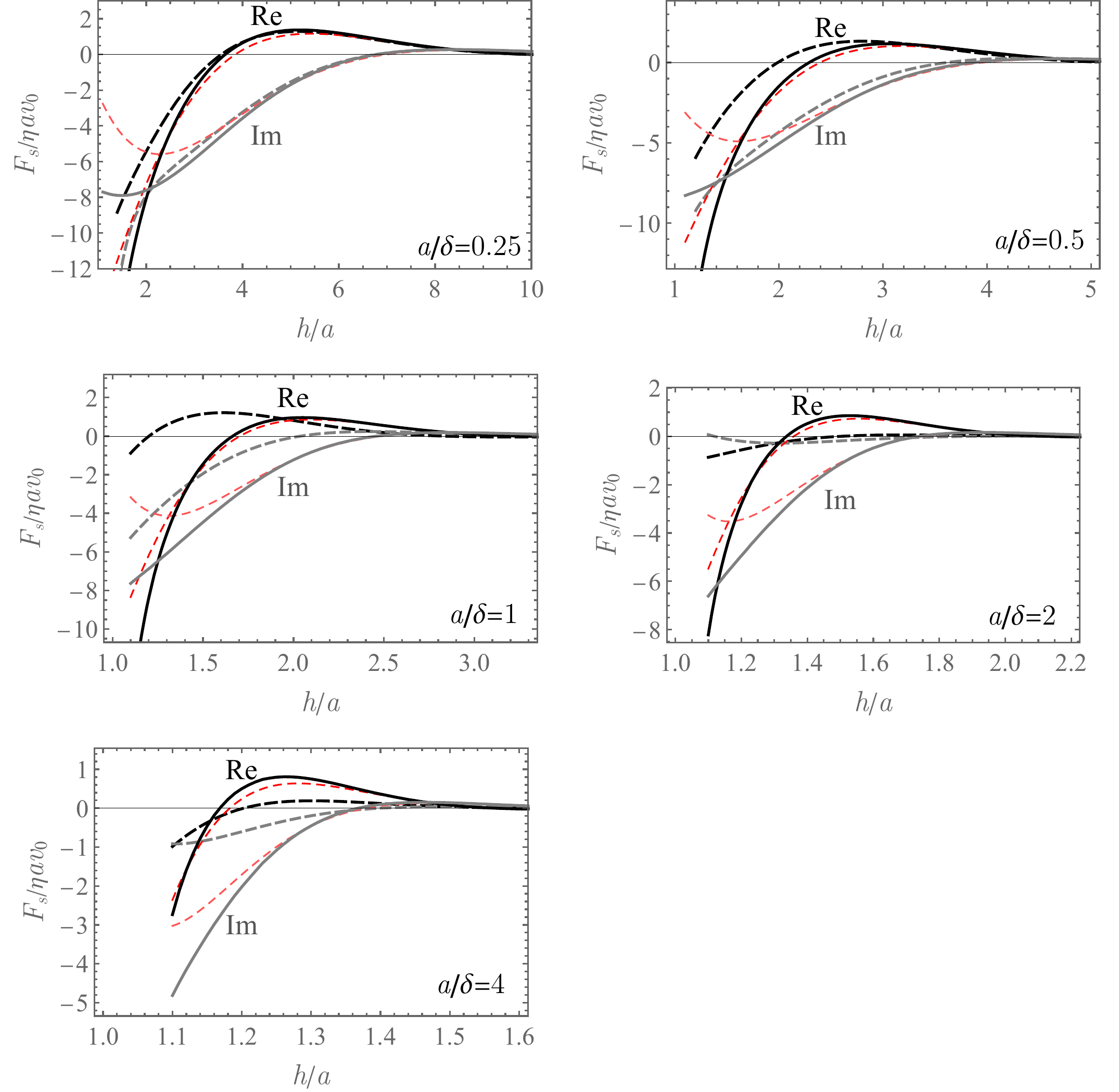}
\caption{Numerically computed real (black solid lines) and imaginary (gray solid lines) part of the complex scaled excess shear force, $F_s/\eta v_0 a$, exerted on the oscillating plate due to a \emph{stationary} particle vs. the scaled separation distance, $h/a$, for several values of $a/\delta$ as indicated by labels. Distant-particle approximation (\ref{Ffasymp}) are shown as thin (red) short-dashed lines, and the small-particle asymptotic result Eq.~(\ref{Ffar}) as thick long-dashed lines. }
\label{fig:Fs}
\end{center}
\end{figure*}
The results for the complex excess shear force, $F_s/\eta v_0 a$ (real and imaginary part, solid lines) exerted on an oscillating plate due to a stationary particle are shown in Fig.~\ref{fig:Fs} vs. the scaled separation distance $h/a$ for different values of $a/\delta$ ranging from $0.25$ to $4$. Numerical results (solid curves) are shown together with the asymptotic small-particle limit (\ref{Ffar}) (bold long-dashed lines) and distant-particle approximation in Eq.~(\ref{Ffasymp}) (thin  short-dashed lines). It can be readily seen that the small particle limit is fairly accurate for a small particle (e.g., $a/\delta=0.25$), and fails for larger particles. In contrast, the distant particle approximation shows an excellent agreement with the numerical results at nearly all separations (besides close proximity) and particle sizes.
\begin{figure*}[tbh]
\begin{center}
\includegraphics[width=0.70\textwidth]{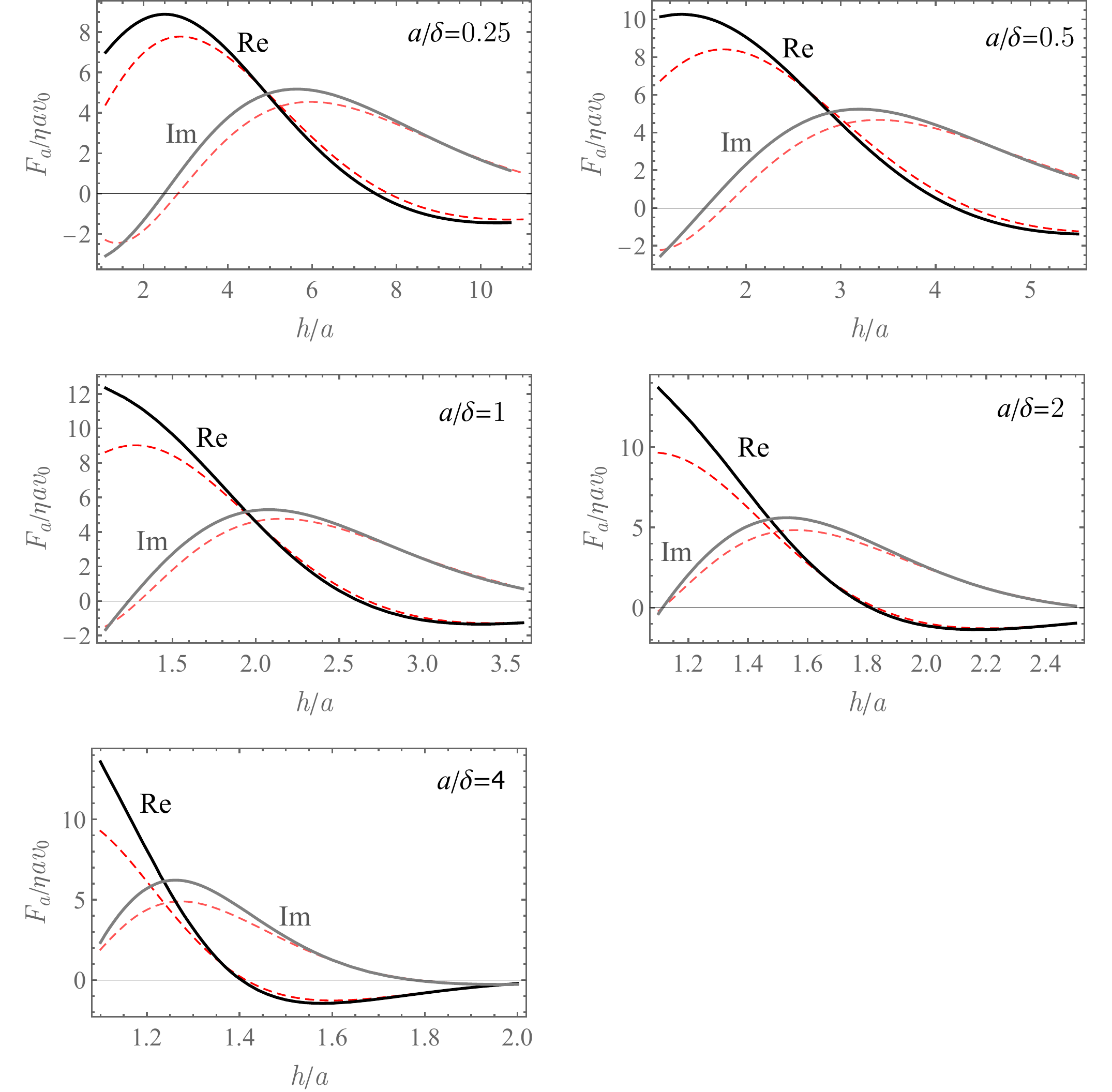}
\caption{Numerically computed real (black solid lines) and imaginary (gray solid lines) part of the complex scaled excess shear force, $F_a/\eta v_0 a$, exerted on the oscillating plate due to an \emph{a firmly adhered} particle vs. the scaled separation distance, $h/a$, for several values of $a/\delta$. Distant-particle approximation in Eq.~(\ref{Faasymp}) are shown as red dashed lines.}
\label{fig:Fa}
\end{center}
\end{figure*}

\begin{figure*}[tbh]
\begin{center}
\includegraphics[width=0.70\textwidth]{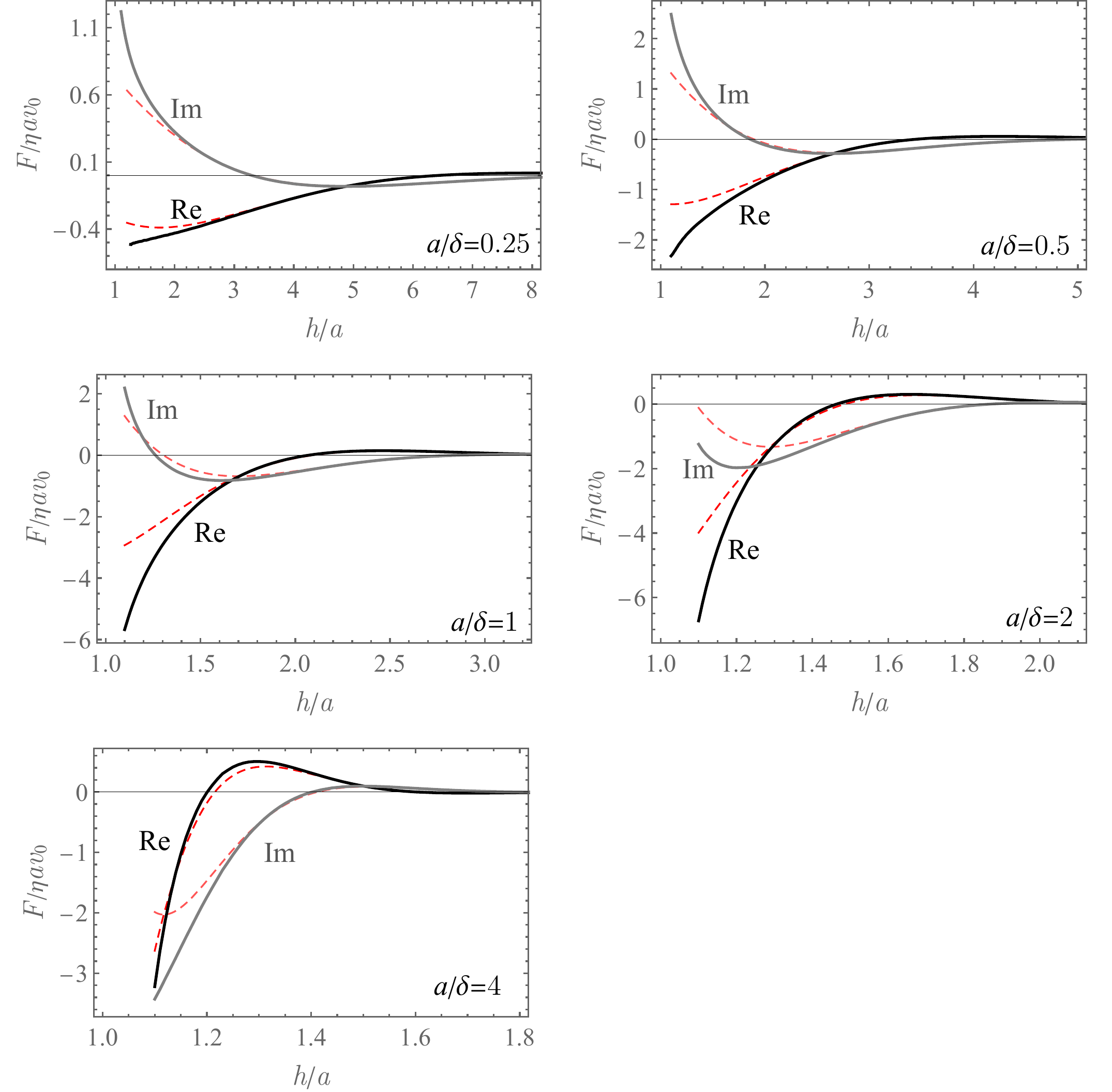}
\caption{Numerically computed real (black solid lines) and imaginary (gray solid lines) part of the complex scaled excess shear force, $F/\eta v_0 a$, exerted on the oscillating plate due to a neutrally buoyant ($\xi=4\pi/3$) \emph{freely suspended} particle vs. the scaled vertical separation distance, $h/a$, for several values of $a/\delta$. Distant-particle approximation in Eq.~(\ref{Fasymp}) are shown as red dashed lines. \label{fig:Ffree}}
\end{center}
\end{figure*}

The analogous results for the excess shear stress due to adsorbed and freely suspended particles are shown in Figs.~\ref{fig:Fa} and Figs.~\ref{fig:Ffree}, respectively. In these figures we only show the comparison between the numerical results (solid lines) and the distant-particle approximations in Eqs.~(\ref{Faasymp}) and (\ref{Fasymp}) (red dashed lines).
As before, the agreement between the numerical results and the approximate solution is quite impressive for all values of $a/\delta$ and for nearly all separation distances, besides very close proximity. Notice that in comparison to a stationary or a freely suspended particle, the adsorbed particle alters the shear force exerted on the plane at longer distances due to fact that it oscillates in-sync with the plane creating a flow disturbance at $z\!=\!h$, while the effect of either stationary  or freely suspended particle is due to reflection of the flow originated at the plate at $z\!=\!0$. The comparison (not shown) of the numerical results with the asymptotic limit of small particle $a\!\ll\!\delta$ in Eq.~(\ref{Fasymp0}) shows a close agreement for $a/\delta\lesssim 0.25$.

\begin{figure*}[tbh]
\begin{center}
\includegraphics[width=0.70\textwidth]{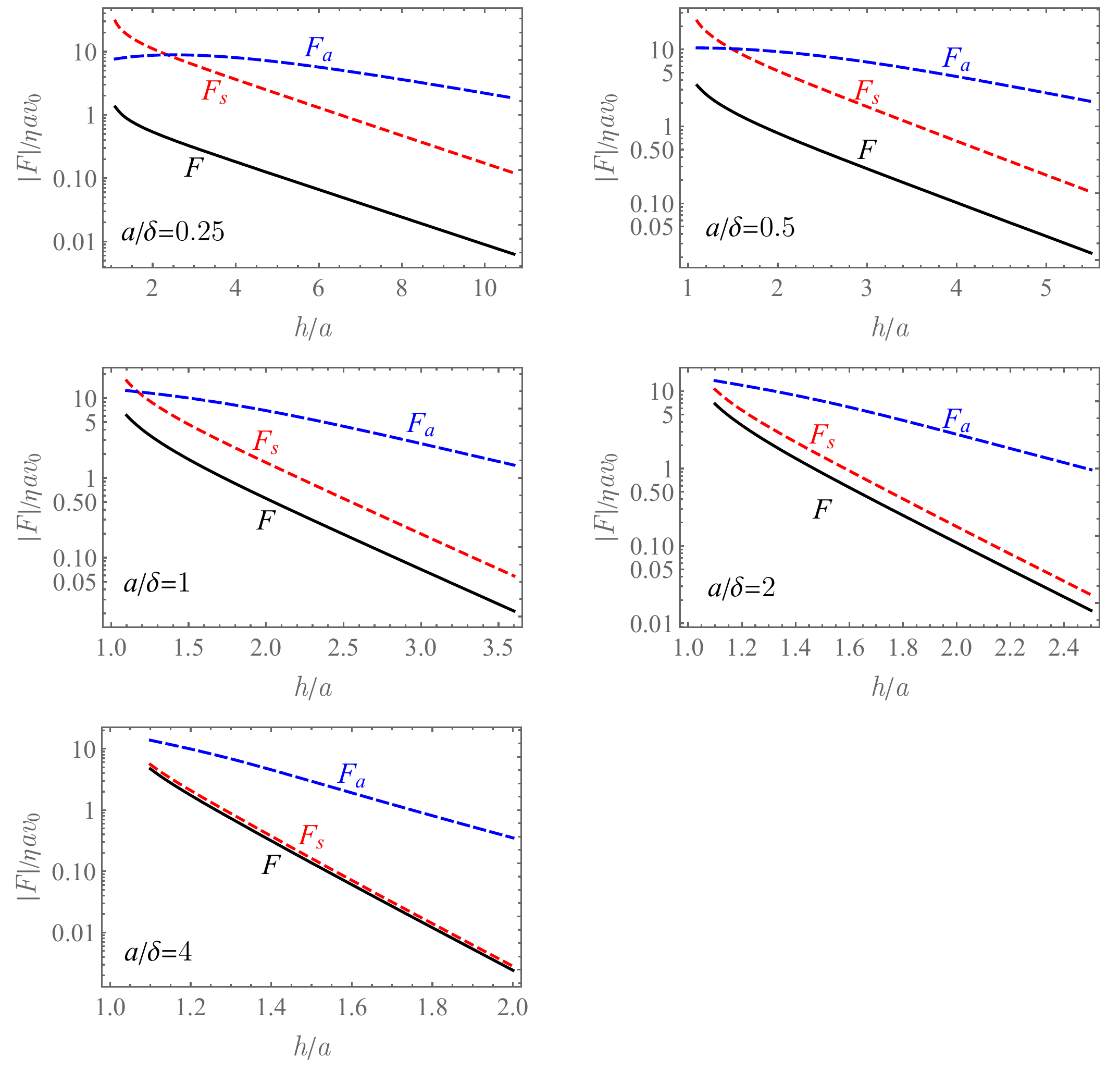}
\caption{Comparison of the numerically computed excess shear force, $|F|/\eta v_0 a$, exerted on the oscillating plate due to a stationary (red short-dashed lines), neutrally buoyant freely suspended (black solid lines) and adsorbed (blue long-dashed line) particles, vs. the separation distance, $h/a$, for several values of $a/\delta$.  \label{fig:FAbs}}
\end{center}
\end{figure*}

To provide a further comparison of the excess shear force due to a stationary, adsorbed and freely suspended (black solid lines) particles, we plot its absolute value computed numerically vs. the separation distance (in log-linear coordinates) for several distinct values of $a/\delta$ (see Fig.~\ref{fig:FAbs}). It can be readily seen that typically $|F|\!<\!|F_s|, |F_a|$, while for larger particles the excess stress due to a freely suspended particle tends to that due to a stationary particle, while $|F|\approx |F_s|\!<\!|F_a|$. The reason for that, is the fact that as the particle becomes larger, the larger is the portion of its surface facing nearly quiescent fluid beyond the penetration depth at $z\!>\!\delta$, leading to suppression of its oscillatory motion driven by the plate oscillations, $V,\, \Omega \rightarrow 0$. Another observation is that the stress perturbation due to a small adsorbed particle is not very sensitive to the proximity to the oscillating plate. For instance, for $a/\delta\!=\!0.25$ the value of $|F_a|/\eta v_0 a$ varies less than $14\%$ for $1\!<\!h/a\!<\!4.5$. The limit of vanishing separations cannot be accurately resolved with our numerical scheme, as it requires solution at exponentially small separation distance, $|\ln \epsilon|\gg 1$, where $\epsilon\!=\!h/a\!-\!1$. We therefore cannot compute the limiting values of $F$ and $F_a$ and test the anticipated relationship in Eq.~(\ref{FvsFa}) (see Sec.~\ref{sec:lubr}). Similarly, $|F_s|$ is expected to (logarithmically) diverge at vanishing proximity, however, such divergence is not apparent in Figs.~\ref{fig:FAbs} as it requires much smaller values of $\epsilon$.

\begin{figure}[tbh]
\includegraphics[scale=0.6]{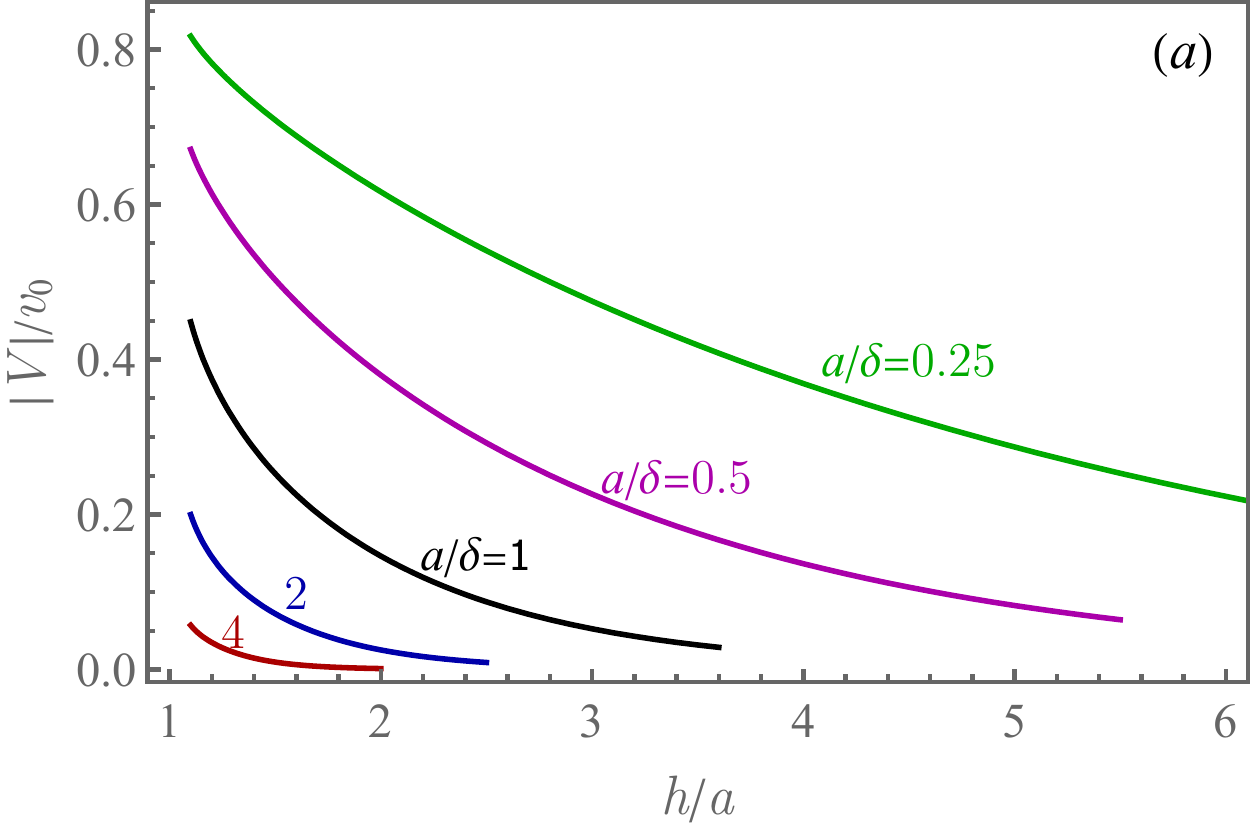} \\
\vspace{3mm}
\includegraphics[scale=0.6]{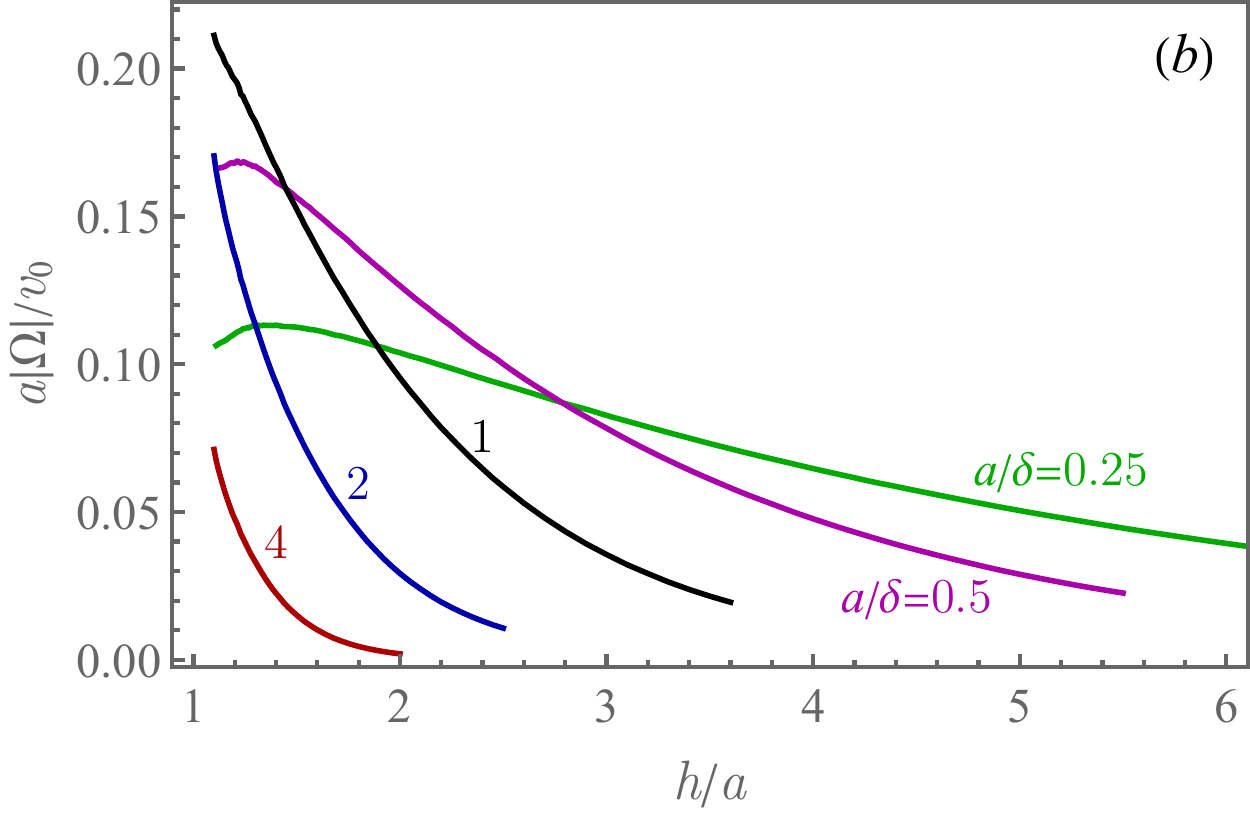}
\caption{Comparison of the numerically computed velocity (absolute value) of a neutrally buoyant freely suspended particle vs. the scaled separation distance, $h/a$; different colors correspond to different values of $a/\delta$ as indicated by the labels. (a) translational velocity, $|V|/v_0$; (b) angular velocity, $a|\Omega|/v_0$. }
\label{fig:VOmega}
\end{figure}

Oscillation of the plate induces oscillatory motion of the freely suspended particles in the fluid above it. The absolute values of the numerically computed translation, $|V|/v_0$, and angular, $a|\Omega|/v_0$, velocities, are depicted in Figs.~\ref{fig:VOmega}a,b , respectively, vs. the separation distance $h/a$ for several values of $a/\delta$ between $0.25$ and $4$. As can be readily seen, the particle translational velocity $|V|/v_0$ decreases monotonically as both $a/\delta$ and $h/a$ increase. As it follows from lubrication approximation (see Sec.~\ref{sec:lubr}) that at close proximity to the plate we have $V\!\to\!1$, while as was mentioned above, this limit applies at exponentially small proximity, $|\ln{\epsilon}|\!\gg\!1$, unaccessible by our numerical scheme.
The particle's angular velocity, $a|\Omega|/v_0$, dependence on distance to the plate, $h/a$, is non-monotonic: it decays at large separations and should vanish at contact, $\Omega \to 0$, as predicted from the lubrication theory (see Sec.~\ref{sec:lubr}). In fact, it can be seen that for small enough particles (e.g., $a/\delta=0.25$) the angular velocity reaches a maximum at some finite distance from the plate, showing the anticipated tendency, $\Omega\!\to\!0$, as $\epsilon\!\to\!0$, while for larger particles the maximum probably moves closer to the plate and more accurate numerical computations are required to test the limiting behavior at vanishing separations.

\begin{figure*}[tbh]
\begin{center}
\includegraphics[scale=0.5]{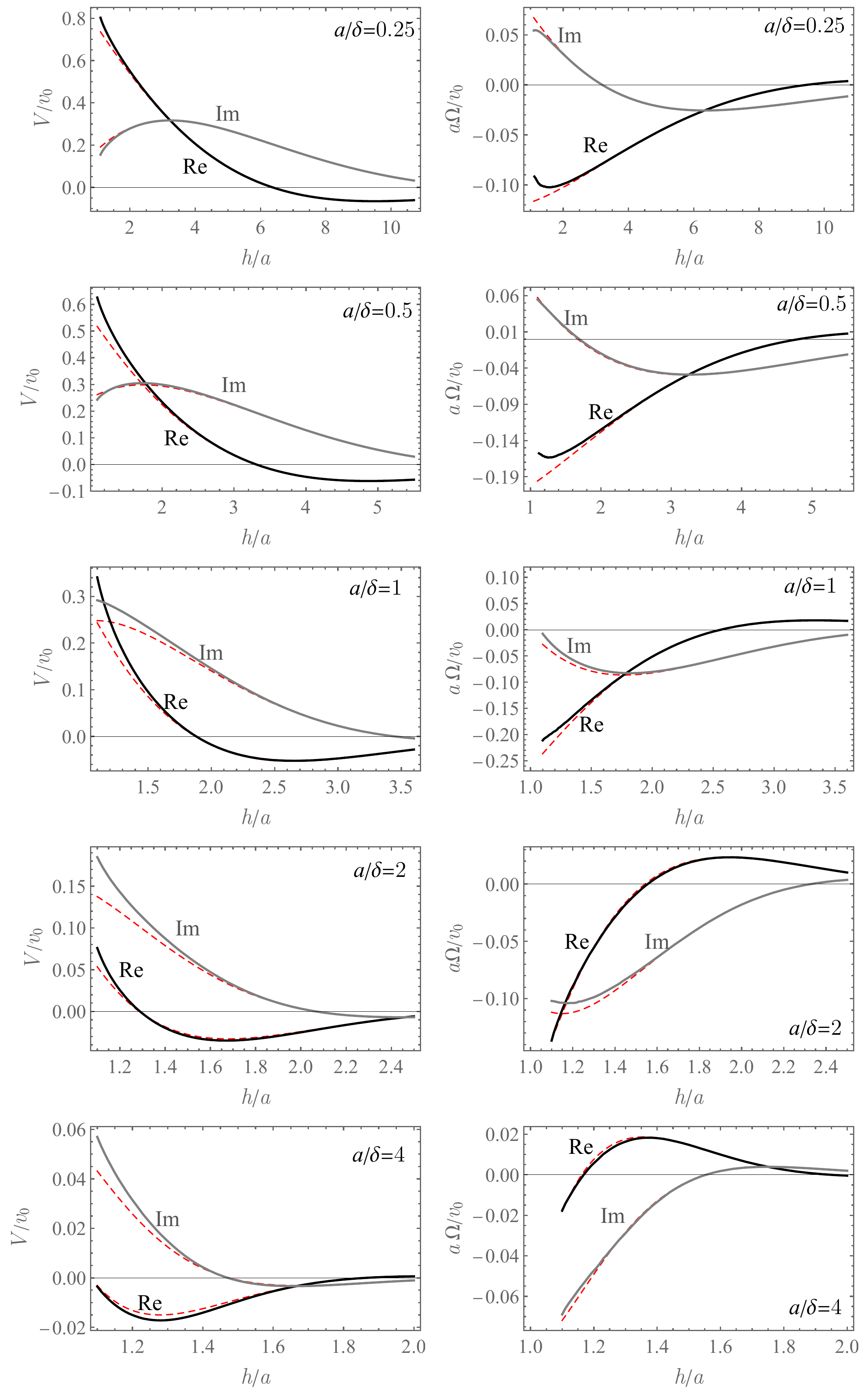}
\caption{ Scaled complex linear $V/v_0$ (left panel) and angular $a\Omega/v_0$ (right panel) velocities of the freely suspended neutrally buoyant ($\xi=4\pi/3$) particle vs. the scaled separation distance, $h/a$, from the oscillating plate for $a/\delta\!=\!0.25$-$4$, as indicated by the labels. Thick solid (black and grey) lines stand for the numerically calculated real and imaginary part, while thin dashed (red) lines correspond to the distant--particle approximation. \label{fig:VOmReIm}}
\end{center}
\end{figure*}

The numerically computed real and imaginary part of the linear, $V/v_0$, and angular, $a\Omega/v_0$, velocities are also depicted in Fig.~\ref{fig:VOmReIm} vs. the distance $h/a$ above the oscillating plate for several values of $a/\delta$ in the range $0.25$--$4$, together with the corresponding predictions of the approximate distant-particle theory in Eqs.~(\ref{Vasymp})--(\ref{Omasymp}). It can be readily seen that the agreement between the numerical results and the approximate theory is excellent for all values of $a/\delta$ in a wide range of separation distances $h/a$ down to close proximity.

\section{Discussion and concluding remarks}

In the present paper we studied the effect of the spherical particle of radius $a$ on the net shear force exerted on horizontal plate fast oscillating in a viscous fluid. The excess shear force depends on three dimensionless parameters: the viscous penetration depth, $\delta/a$ (or, alternatively, the Roshko number, $\mathrm{Ro}\!=\!a^2\omega/\nu$) that characterizes the fluid inertia, the parameter $\xi$ (or, alternatively, the Stokes number, $\mathrm{St}=\mathrm{Ro}\xi$) that measures the particle inertia, and the scaled separation distance $h/a$.

We first considered the simplest (auxiliary) case of a stationary particle, which was exploited to study more complex cases of  freely-suspended and adsorbed particle (i.e., oscillating in-sync with the plate as a whole). The study combines asymptotic analysis corresponding to a distant particle and rigorous numerical simulations (using FEM) valid at arbitrary proximity of the particle to the oscillating plate. We demonstrated that the limit of a distant particle gives accurate predictions well beyond its formal domain of validity. In fact, it provides a surprisingly accurate estimate of the net excess shear force down to very small separation distances. The reason for such accuracy is that unsteady Stokes flow with a given frequency generates a very specific flow, given by the sum of $-\nabla p/\lambda^2$ and the flow that decays exponentially fast away from the particle's surface \cite{fouxon22}. The long-range component of the flow $-\nabla p/\lambda^2$ exerts zero shear force on the plane. Therefore the force is controlled by the exponentially decaying flow component and its fast decay results in surprising accuracy of the theory even at close proximity.

The anticipated hydrodynamic contribution to the QCM-D impedance could be significant, in particular for small particles (small values of $a/\delta$). The dimensionless purely ``inertial" (i.e., due to solid inertia) impedance due to a layer of adsorbed particles with surface density $\tilde n=N/A$ is  given by the Sauerbrey equation, $\widehat{\mathcal Z}_S=\ri m_e\omega/(\eta \tilde n a)=-4\pi (\rho_e/\rho) \lambda^2/3$, where $\rho_e\!=\!\rho_s\!-\!\rho>0$ is the excess density and $\lambda^2=-2\ri (a/\delta)^2$. $\widehat{\mathcal Z}_S$ corresponds to negative frequency shift, $\Delta f \propto \rRe[\ri \mathcal Z_S]<0$ and zero dissipation $\Delta\Gamma \propto \rIm[\ri \mathcal Z_S]=0$. On the other hand, the hydrodynamic contribution to the impedance due to dilute dispersion of particles with radii $a$ can be found by summing up individual contributions from ${\mathcal Z}=v_c^{-1} \int n F(h) dh$, where $F(h)$ is the derived excess shear force and $n$ is (height-dependent) particle number density. (The effect of polydispersity can be similarly taken into account upon averaging over the particle size-distribution.) Using the simple result (\ref{Fasymp0}) we obtained for small particles ($a\!\ll\!\delta$) and assuming constant particle bulk density $n$, we readily find upon integrating over $h$ the dimensionless hydrodynamic impedance $\widehat{\mathcal Z}={\mathcal Z}/(\eta n a^2)\!\approx \!-2\pi\lambda\re^{-2\lambda} (1+\frac{2\rho_s}{3\rho})$. Substituting $a/\delta=0.1$ and using polystyrene particles in water with $\rho_s/\rho=1.05$, we readily find that the expected value of the scaled inertial impedance is $\widehat{\mathcal Z}_S\approx 0.026\,\ri$. The hydrodynamic counterpart gives $\widehat{\mathcal Z}\approx -0.52+0.34\,\ri$. Comparing the imaginary part of the two contributions we obtain $|\rIm[\widehat{\mathcal Z}]/\rIm[\widehat{\mathcal Z}_S]|\sim 80$. In other words, for weakly adhering small nanoparticles with surface density controlled by the bulk volumetric concentration, $\tilde n\sim n a$, the hydrodynamic contribution will dominate the QCM-D signal. Even for high surface density ${\tilde n}/(na) \!\sim\! 10^3$ the hydrodynamic part would yet contribute about 10\% of the signal. The hydrodynamic signature of small particles that are adsorbed at the resonator is even more pronounced. Using the small-$\lambda$ limit of Eq.~(\ref{Faasymp0}) at contact we find $\widehat{\mathcal Z}_a={\mathcal Z}_a/(\tilde n \eta a)\approx 6\pi\lambda$. Thus for $a/\delta=0.1$ we have $\widehat{\mathcal Z}_a\approx 1.9(1\!-\!\ri)$, and the ratio of the hydrodynamic contribution to the inertial (Sauerbrey) impedance yields $|\rIm[\widehat{\mathcal Z}_a]/\rIm[\widehat{\mathcal Z}_S]|\sim 450$. Even for heavy particles with $\rho_s/\rho\!=\!1.9$ we still find $|\rIm[\widehat{\mathcal Z}_a]/\rIm[\widehat{\mathcal Z}_S]|\sim 25$.

One of the major results of the present paper is the analysis of vanishing separation between the particle and the oscillating plate. We showed that the standard lubrication theory developed for modeling near-contact particle-particle or particle-wall interactions in \emph{steady} Stokes flows \cite{kim}, can be extended to unsteady Stokes equations. Besides the standard requirement of the lubrication theory, stating that the clearance between the particle and the plate must be much smaller than the particle's radius, $\epsilon\!=\!h/a\!-\!1\ll 1$, there is an extra requirement, $\epsilon \!\ll\! \delta/a$, that comes from unsteadiness and which imposes strict limitations on $\epsilon$ for large particles. Applying the lubrication theory, we demonstrated that in accord with the \emph{ad hoc} assumption in Ref.~\cite{Busca21}, in the limit $h \to a$ the freely-suspended particle hydrodynamically ``adheres" to the oscillating plate, as its translation velocity tends to that of the plate and its angular velocity vanishes. However, such adherence does not warrant the equality of the excess shear force due to a freely suspended and the adsorbed particle, as was suggested in Ref.~\cite{Busca21}, as the difference between the limiting values of $F$ and $F_a$ at contact depends on $\lambda$. The accurate limiting behavior of the excess stress at contact would require more accurate numerical modeling and will be conducted elsewhere. Notice also that our analysis also does not take into account the non-hydrodynamic short-range forces acting on the particle at close proximity.

Determining the shear force exerted on the oscillating plate in presence of a freely suspended particle with finite inertia requires the solution of the coupled system of equations for the flow and fluid-mediated time-periodic translational and rotational motions of the particle.
The oscillatory particle motion determined numerically and analytically (using distant particle approximation) can be further used in calculation of the nonlinear vertical force exerted on a freely suspended particle above an oscillating plate. Such force is due to the nonlinear terms in the Navier-Stokes equations. Although for small-amplitude oscillations the non-linearity is typically small, it leads to the emergence of the finite constant vertical force  $\sim \rho v_0^2 a^2$ exerted on the particle and which is absent when considering linear unsteady Stokes equations. Such nonlinear vertical force is of interest, as it can alter deposition rate of small substances at the QCM-d resonator. We have recently determined the nonlinear vertical force exerted on a \emph{stationary} particle \cite{prl20}, which may also be applicable to freely suspended \emph{heavy} inettial particles. In particular, we found that in the limit of vanishing separation, large particles would experience the lift force, directed \emph{away} from the resonator \cite{prl20}. On the other hand, it was theoretically predicted that for neutrally buoyant and freely suspended particles the vertical force is directed \emph{towards} the resonator (i.e., anti-lift) \cite{lj13}, indicating its strong dependence on particle's inertia. Notice, however, that the calculations in Ref.~\cite{lj13} are approximate, as they assumed that the particle follows the motion of the undisturbed background flow that would persist in the absence of the particle. Since in the present paper the motion of the freely suspended particle is determined self-consistently for an arbitrary $a$ and $\mathrm{St}$, it paves a way to calculating the nonlinear vertical force rigorously. Such study, however, requires going beyond the linear Stokes flow approximation and shall be considered elsewhere.

In accord with the approximate theory in Ref.~\cite{Busca21}, our work confirms the notion that quantitative analysis of QCM-D measurements calls for accurate account of hydrodynamics of the adjacent fluid altered by presence of the suspended small substances. The accurate account of hydrodynamics is complex due to nontrivial geometry and unsteadiness. This work demonstrates, however, that analytical progress is possible. The next step could be introduction of non-hydrodynamic (e.g., adhesion forces) that may compete with hydrodynamics at close proximity and alter the impedance reading.

\section*{Acknowledgements}
This work was supported, in part, by the Israel Science Foundation (ISF) via the grant No. 1744/17 (A.M.L.)

\newpage
\begin{appendices}
\appendix

\section{Derivation by Schofield and Delgado-Buscalioni \cite{Busca21} revisited} \label{app:A}
\setcounter{equation}{0}
\renewcommand{\theequation}{A\arabic{equation}}

In this Appendix we revisit the theory described in Ref.~\cite{Busca21} for the solution of Eqs.~(\ref{fond}) resulting in approximate closed-form expressions for the excess shear force (or the hydrodynamic impedance) due to particle in the close proximity to the oscillating plate. This approach involved a number of \emph{ad hoc} approximations and simplifying assumptions that we aim to re-examine. The authors made use of the induced force representation introduced in Ref.~\cite{ma74}. Within this representation, the (no slip) boundary condition at the particle surface is provided by an the volumetric force distribution within the particle. The flow then persists everywhere, including the particle's interior and it can be represented by a volume integral of the Green's function. Then an \emph{ad hoc} ansatz for the induced force density assuming uniform flow disturbance within the particle is used. This ansatz yields an approximate boundary integral representation of the flow, similar (but not identical) to the exact formulation (see, e.g., \cite{pzr,fl18}). Each portion of the particle surface generates flow which is proportional to the Green's function times a coordinate-dependent factor ${\tilde f}$. Then using the expression for the shear stress exerted at the plate due to a point force (via the Green's function derived in \cite{Felderhof2012} and \cite{fl18}), reduces the calculation of the excess shear force (or impedance) to the integral over ${\tilde f}$. Since ${\tilde f}$ involves the unknown particle velocity, excess shear force was determined under the assumption that it undergoes oscillatory translations with the velocity of the plate without rotation, as discussed in the main text (see Sec.~\ref{sec:intro}). The approach of Ref.~\cite{Busca21} leads to the approximate analytical expression for the excess shear force for a particle near contact, which is identical (up to a factor of $\pi$) to the partial result for an absorbed particle, $F_a$, derived in this paper [the last two terms in Eq.~(\ref{fa})].

We consider the same decomposition $\bm v=\re^{-\lambda z}\hat{\bm x}+\bm u$ that was used in Sec.~\ref{stationary}. The flow perturbation due to the presence of a particle, $\bm u$, satisfies:
\begin{eqnarray}&&\!\!\!\!\!\!\!\!\!\!\!\!\!
\lambda^2 \bm u\!=\!-\nabla
 p\!+\!\nabla^2\bm u,\ \ \nabla\cdot\bm u\!=\!0,
\nonumber\\&&\!\!\!\!\!\!\!\!\!\!\!\!\! \bm u(z=0)= 0,\ \ \bm u(r=1)=V \hat{\bm x}+\Omega\, \hat{\bm y}\times \bm r-\re^{-\lambda z}\hat{\bm x}, \label{bcu}
\end{eqnarray}
cf. Eq.~(\ref{fon17}). The solution obeys the boundary integral representation \cite{pzr,fl18}
\begin{eqnarray}
u_{i}(\bm x)&=&-\oint_{|\bm x'-\bm x_c|=1} \frac{G_{ik}(\bm x, \bm x')\sigma_{kl}(\bm x')dS'_l}{8\pi} \nonumber\\
&& \!+\!\oint_{|\bm x'-\bm x_c|=1} u_{k}(\bm x')\sigma^i_{kl}(\bm x', \bm x) dS'_l\,, \label{repr}
\end{eqnarray}
where $\sigma_{kl}$ is the stress tensor associated with the flow $\bm u$ and integration is performed over $\bm x'$. The Green's function is defined via the solution of the following problem:
\begin{eqnarray}&&\!\!\!\!\!\!\!\!\!\!\!\!\!
-\nabla p^k\!+\! \nabla^2\bm u^k-\lambda^2\bm u^k=-{\hat x}_k\delta(\bm x-\bm x'),\ \ \nabla\cdot\bm u^k\!=\!0, \nonumber
 \\&&\!\!\!\!\!\!\!\!\!\!\!\!\! \bm u^k(z\!=\!0)\!=\!\bm u^k(r\to\infty)\!=\!0\,,\label{uns}
\end{eqnarray}
where ${\hat x}_k$ is unit vector along $k$th axis. The solution $u^k_i(\bm x, \bm x')$, whose dependence on force parameters $k$ and $\bm x'$ is made explicit, defines $G_{ik}(\bm x, \bm x')$ as,
\begin{eqnarray}&&\!\!\!\!\!\!\!\!\!\!\!\!\!
u^k_i(\bm x, \bm x')=\frac{G_{ik}(\bm x, \bm x')}{8\pi}.
\end{eqnarray}
Notice that the factor $8\pi$ is missing in the denominator of the last equation for $G_{ik}(\bm x, \bm x')$ in Ref.~\cite{Busca21}. Finally $\sigma^i_{kl}$ in Eq.~(\ref{repr}) is the stress tensor of the flow $\bm u^k$. Application of the divergence theorem and
$\lambda^2 u^k_i=\partial_l\sigma^i_{kl}$ gives \cite{fl18}
\begin{eqnarray}&&\!\!\!\!\!\!\!\!\!\!\!\!\!
\oint_{|\bm x'-\bm x_c|=1}\!\! \sigma^i_{kl}(\bm x', \bm x) dS'_l\!=\!\lambda^2 \int_{|\bm x'-\bm x_c|<1}\!\! \frac{G_{ik}(\bm x, \bm x')}{8\pi} d\bm x'.
\end{eqnarray}
Similarly using the divergence theorem and symmetry of $\sigma^i_{kp}$ in the lower indices we have
\begin{eqnarray}&&\!\!\!\!\!\!\!\!\!\!\!\!\!
\oint_{|\bm x'-\bm x_c|=1}\!\! \epsilon_{kyp}r'_p \sigma^i_{kl}(\bm x', \bm x) dS'_l
\nonumber\\&&\!\!\!\!\!\!\!\!\!\!\!\!\!
=\lambda^2 \int_{|\bm x'-\bm x_c|<1}\!\! \frac{\epsilon_{kyp}r'_pG_{ik}(\bm x, \bm x')}{8\pi} d\bm x',
\end{eqnarray}
where $\bm r'=\bm x'-\bm x_c$.
Finally, we also have by the divergence theorem
\begin{eqnarray}&&
\oint_{|\bm x'-\bm x_c|=1}\!\! \re^{-\lambda z'}\sigma^i_{xl}(\bm x', \bm x) dS'_l=
\lambda^2\int_{|\bm x'-\bm x_c|<1}\!\! \frac{G_{ix}(\bm x, \bm x')}{8\pi}
\nonumber\\&&
\re^{-\lambda z'} d\bm x'-\lambda \int_{|\bm x'-\bm x_c|<1}\re^{-\lambda z'}\sigma^i_{xz}(\bm x', \bm x)d\bm x', \label{str}
\end{eqnarray}
where we used the symmetry $G_{ik}(\bm x, \bm x')=G_{ki}(\bm x', \bm x)$ (see, e.g., \cite{fl18}).

Substituting the boundary conditions in Eq.~(\ref{bcu}) into Eq.~(\ref{repr}) and using the above relations gives
\begin{eqnarray}
u_{i}(\bm x)&=&-\!\oint_{|\bm x'-\bm x_c|=1}\!\!\!\!\!\! \frac{G_{ik}(\bm x, \bm x')\sigma_{kl}(\bm x')dS'_l}{8\pi} \nonumber\\
&& +\lambda^2 \!\!\int_{|\bm x'-\bm x_c|<1}\!\!\!\!\!\!\left(V \delta_{kx}+\Omega \epsilon_{kyp}r'_p\right)\frac{G_{ik}(\bm x, \bm x')}{8\pi} d\bm x' \nonumber\\
&& -\lambda^2\int_{|\bm x'-\bm x_c|<1}\!\! \re^{-\lambda z'} \frac{G_{ix}(\bm x, \bm x')}{8\pi} d\bm x' \nonumber\\
&& +\lambda \int_{|\bm x'-\bm x_c|<1}\re^{-\lambda z'}\sigma^i_{xz}(\bm x', \bm x)d\bm x'. \label{resp}
\end{eqnarray}
The above flow representation is rigorous and does not involve any approximations. The approximate flow representation derived in \cite{Busca21} upon including the $8\pi$ factor in the definition of the Green's function and re-scaling the dimensional variables, reads
\begin{eqnarray}&&\!\!\!\!\!\!\!\!\!\!\!\!\!
u_{i}(\bm x)=-\oint_{|\bm x'-\bm x_c|=1} \frac{G_{ik}(\bm x, \bm x')\sigma_{kl}(\bm x')dS'_l}{8\pi}
\nonumber\\&&\!\!\!\!\!\!\!\!\!\!\!\!\!
+\lambda^2 \left(V-\frac{3}{4\pi}\int \re^{-\lambda z}d\bm x\right)\int_{|\bm x'-\bm x_c|<1}\frac{G_{ix}(\bm x, \bm x') d\bm x'}{8\pi}.\label{aresp}
\end{eqnarray}
The comparison of the approximate (\ref{aresp}) and the exact (\ref{resp}) representations reveal three discrepancies. First, the approximate representation (\ref{aresp}) does not include the term $\propto \Omega$, as Ref.~\cite{Busca21} assumed \emph{ad hoc} that at vanishing separations the particle angular velocity $\Omega$ tends to zero, and therefore it does not contribute to the excess shear force and impedance to the leading approximation. As we demonstrated in Sec.~\ref{sec:lubr}, unless $a\gg \delta$, the (vanishing) angular velocity does contribute to the excess shear exerted on the plate at the leading order. Therefore, the term $\propto \Omega$ in the exact integral representation (\ref{resp}) cannot be omitted in general. Secondly, the last term in Eq.~(\ref{resp}) is missing in the approximate representation (\ref{aresp}). Finally, (\ref{aresp}) assumes that
\begin{eqnarray}&&
\int_{|\bm x'-\bm x_c|<1}\!\! \re^{-\lambda z'} \frac{G_{ix}(\bm x, \bm x')d\bm x'}{8\pi}\approx 
\nonumber\\&&
\frac{3}{4\pi}\int \re^{-\lambda z'}d\bm x' \int_{|\bm x'-\bm x_c|<1}\!\!  \frac{G_{ix}(\bm x, \bm x')d\bm x'}{8\pi}.
\end{eqnarray}
It is readily seen from Eq.~(\ref{str}) that the last two approximations hold given that the variation of $\re^{-\lambda z}$ within the sphere can be neglected. This gives the condition of validity of the approximate representation of \cite{Busca21} as $\delta \gg a$. Similarly, anther approximations used by \cite{Busca21}, such as point-particle representation, assumes that the particle is small. The comparison to the numerical results in Ref.~\cite{Busca21} suggested that the approximation is accurate for $a/\delta\lesssim 2$, well beyond $a/\delta\ll 1$, indicating that in the case of neutrally buoyant adsorbed particles, the terms that could result in deviation between Eqs.~(\ref{resp}) and Eq.~(\ref{aresp}), say at $a\!\approx \!\delta$, are probably small.

\section{Calculation of $C_X$ \label{app:B}}
\setcounter{equation}{0}
\renewcommand{\theequation}{B\arabic{equation}}

The integral for $C_X$ in Eq.~(\ref{ifd}) can be re-written as:
\begin{eqnarray}&&
C_X\equiv \int \! \left[\partial_y \left(\re^{\lambda (1-r)}\frac{zX}{r}\right)\!-\!\partial_z \left(\re^{\lambda (1-r)}\frac{yX}{r}\right)\right] dxdy
\nonumber \\&&
=-\re^{\lambda}\frac{\partial (h^2 \mathcal{G})}{\partial h},\ \ \mathcal{G}\equiv -\frac{1}{h^2}\int_{z=-h} \frac{ yX}{r} \re^{-\lambda r} dxdy, \label{id}
\end{eqnarray}
where the dimensionless integral ${\mathcal G}$ can be calculated in spherical coordinates. We notice that for any function $g$ the plane integral satisfies
\begin{eqnarray}&&\!\!\!\!\!
\int_{z=-h}\! g dxdy\!=\!\int \! g \delta(z\!+\!h) d\bm x\!=\!\int \! g \delta(r\cos{\theta}\!+\!h) r^2\sin{\theta} dr
\nonumber\\&&\!\!\!\!\!
\times d\theta d\phi =-h^2\int \!d\phi\int_{\pi/2}^{\pi}  \frac{\tan{\theta}}{\cos^2{\theta}} g|_{r=-h/\cos{\theta}}\, d\theta,\label{transofm}
\end{eqnarray}
where $\delta(x)$ stands for the Dirac $\delta$ function (not to be confused with the penetration depth). The integral over the azimuthal angle in the resulting expression for ${\mathcal G}$ involves $\int \! \sin\phi Y_{lm}d\phi$ which is non-zero only if $|m|=1$. Using
\begin{eqnarray}&&
Y_{l1}=\sqrt{\frac{(2l+1)}{4\pi l(l+1)}}P_l^1(\cos{\theta})\re^{\ri\phi}=-Y_{l,-1}^*,\label{sd}
\end{eqnarray}
we have
\begin{eqnarray}&&
\int \! \sin{\phi} Y_{lm}d\phi= \ri\pi \sqrt{\frac{(2l+1)}{4\pi l(l+1)}}P_l^1(\cos{\theta})\left(\delta_{m1}+\delta_{m, -1}\right).\nonumber
\end{eqnarray}
Thus, using Eq.~(\ref{X}) we find that ${\mathcal G}$ defined in (\ref{id}) has the form
\begin{eqnarray}&&
{\mathcal G}\!=\!\ri \sum_{l }\left({\tilde c}_{l1}+{\tilde c}_{l, -1}\right)\sqrt{\frac{(2l+1)\pi}{4 l(l+1)}}
 \label{ian} \\&&
\int_{\pi/2}^{\pi}\!\!P_l^1(\cos\theta)\kappa_{l}\left(-\frac{\cos\theta}{\lambda h}\right) \exp{\left(\frac{\lambda h}{\cos\theta}\right)} \frac{\tan^2\theta}{\cos\theta} d\theta.\nonumber
\end{eqnarray}
This formula simplifies in the considered limit of $|\lambda|h\gg 1$ where it is determined by a narrow vicinity of point on the plane $\theta=\pi$ which is closest to the sphere. The expansion of the exponential
\begin{eqnarray}&&\!\!\!\!\!\!\!\!\!\!\!\!\!
\exp\left(\frac{\lambda h}{\cos\theta}\right)\approx \exp\left(-\lambda h -\lambda h\frac{(\pi-\theta)^2}{2}\right),
\end{eqnarray}
demonstrates that the effective domain of integration over $\theta$ is $|\pi-\theta|\sim \left|\lambda h\right|^{-1/2}\ll 1$.  Therefore we can approximate the integrand by the leading order term at small $\pi-\theta$. We use that
\begin{eqnarray}&&\!\!\!\!\!\!\!\!
P_l^1(x)=\sqrt{1-x^2}f_l(x),\ \ f_l(x)\equiv -\frac{1}{2^l l!}\frac{d^{l+1}}{dx^{l+1}}(x^2-1)^l ,\nonumber
\end{eqnarray}
where $f_l(-1)$ is finite and given by
\begin{eqnarray}&&\!\!\!\!\!\!\!\!
f_l(-1)=\frac{(-1)^l l(l+1)}{2}. \label{flor}
\end{eqnarray}
We find that Eq.~(\ref{ian}) becomes
\begin{eqnarray}&&
{\mathcal G}\!\approx \!- \ri \re^{-\lambda h} \sum_{l }\left({\tilde c}_{l1}+{\tilde c}_{l, -1}\right) \sqrt{\frac{(2l+1)\pi}{4 l(l+1)}}f_l(-1)
\nonumber \\&&
\kappa_{l}\left(\frac{1}{\lambda h}\right) \int_{\pi/2}^{\pi}\!\!(\pi-\theta)^3\exp\left(-\frac{\lambda h(\pi-\theta)^2}{2}\right)d\theta. \label{ian1}
\end{eqnarray}
Using $\int_0^{\infty} t^3 \re^{-t^2}dt=1/2$, the last integral in (\ref{ian1}) reduces to $2/(\lambda h)^2$. Using $\kappa_l\approx 1$, cf. Eq.~(\ref{mod}), we conclude that
\begin{eqnarray}&&\!\!\!\!\!\!\!\!
C_X\!=\!- \ri f_l(-1) \re^{\lambda (1-h)} \sum_{l }\frac{{\tilde c}_{l1}\!+\!{\tilde c}_{l, -1}}{\lambda} \sqrt{\frac{(2l\!+\!1)\pi}{ l(l\!+\!1)}}.
\end{eqnarray}
In order to calculate ${\tilde c}_{l1}+{\tilde c}_{l, -1}$, we notice that $\nabla\times(\re^{-\lambda z}{\hat x})=-\re^{-\lambda z} \lambda {\hat y}$ so that
\begin{eqnarray}&&
{\tilde c}_{l1}+{\tilde c}_{l, -1}=\frac{2 \ri \lambda}{l(l+1)\kappa_l(\lambda^{-1}) } \sqrt{\frac{(2l+1)}{4\pi l(l+1)}}
\\&&
 \int_{r=1} P_l^1(\cos\theta) \sin^2\phi  \sin\theta\, \re^{-\lambda \cos{\theta}}  d\Omega,\nonumber
\end{eqnarray}
see Eqs.~(\ref{cd}) and (\ref{sd}). This gives
\begin{eqnarray}&&
C_X=\re^{\lambda (1-h)} \sum_{l }\frac{\pi (2l+1)f_l(-1)}{ l^2(l+1)^2\kappa_l(\lambda^{-1})}\nonumber \\&&
\times \int_{-1}^1 P_l^1(x) \sqrt{1-x^2} \re^{-\lambda x}  dx. \label{cx1}
\end{eqnarray}
We have for the last integral in (\ref{cx1})
\begin{eqnarray}&&\!\!\!\!\!\!\!\!\!\!\!\!
\int_{-1}^1\re^{-\lambda x} \sqrt{1-x^2}P_l^{1}(x)dx
\nonumber\\&&\!\!\!\!\!\!\!\!\!\!\!\!
= \int_{-1}^1 \re^{-\lambda x}(\lambda x^2-2x-\lambda )P_l(x)dx
\nonumber\\&&\!\!\!\!\!\!\!\!\!\!\!\!
= \left(\lambda \frac{d^2}{d\lambda^2}+2\frac{d}{d\lambda}-\lambda \right)\int_{-1}^1 \re^{-\lambda x} P_l(x)dx. \label{aux0}
\end{eqnarray}
where we used
\begin{eqnarray}&&\!\!\!\!\!\!\!\!\!\!\!\!
\sqrt{1-x^2}P_l^{1}(x)=\frac{x^2-1}{2^l l!}\frac{d^{l+1}}{dx^{l+1}}(x^2-1)^l.
\end{eqnarray}
The last integral in (\ref{aux0}) can be calculated in terms of the modified Bessel functions of half-integer order \cite{prudnikov}:
\begin{eqnarray}&&\!\!\!\!\!\!\!\!\!\!\!\!
\int_{-1}^1\re^{-\lambda x} P_l(x)dx=(-1)^l\sqrt{\frac{2\pi}{\lambda}} I_{l+1/2}(\lambda).
\end{eqnarray}
Further using the Bessel equation for the $I_{l+1/2}$ functions,
\begin{eqnarray}&&\!\!\!\!\!\!\!\!\!\!\!\!
\left[\frac{d^2}{d\lambda^2}+\frac{2}{\lambda}\frac{d}{d\lambda}-\left(1+\frac{l(l+1)}{\lambda^2}\right)\right]
\frac{I_{l+1/2}(\lambda)}{\sqrt{\lambda}}=0,
\end{eqnarray}
the integral in (\ref{aux0}) finally gives,
\begin{eqnarray}&&\!\!\!\!\!\!\!\!\!\!\!\!
\int_{-1}^1\!\!\!\!\re^{-\lambda x}\sqrt{1\!-\!x^2}P_l^{1}(x)dx
\!=\nonumber \\
&& \!\sqrt{\frac{2\pi}{\lambda^3}}(-1)^l l(l\!+\!1)I_{l+1/2}(\lambda).\label{tbl}
\end{eqnarray}
Substituting the last result into Eq.~(\ref{cx1}) and using Eq.~(\ref{flor}), we find that
\begin{eqnarray}&&
C_X\!=\!
\frac{2 \pi^2 }{\lambda^2} \re^{-\lambda h} \sum_{l=1}^{\infty}\frac{(2l\!+\!1)I_{l+1/2}(\lambda)}{K_{l+1/2}(\lambda) },
\label{cx2}
\end{eqnarray}
where we re-wrote polynomials $\kappa_l$ in terms of the modified Bessel functions $K_{l+1/2}$ by using Eq.~(\ref{mod}).

\section{Calculation of $C^H$ \label{app:C}}
\setcounter{equation}{0}
\renewcommand{\theequation}{C\arabic{equation}}

Let us consider the coefficient $C^H$ defined in Eq.~(\ref{ifq}). The contribution of the gradient term in $\bm u^H$ yields zero contribution as integral of a derivative and, therefore, we have
\begin{eqnarray}&&
C^H\!\equiv \!-\lambda \re^{\lambda}\sum_{l m} \int_{z=-h}\!
\re^{- \lambda r}  {\tilde c}^{r}_{lm} \sin{\theta}\cos{\phi}\, Y_{lm}
\nonumber\\&&
\times\left[\frac{4l^2\!-\!1}{2\lambda r} \kappa_{l-1}\left(\frac{1}{\lambda r}\right)\!+\!
\kappa_{l-2}\left(\frac{1}{\lambda r}\right) \right] dxdy\,.
\end{eqnarray}
By using Eq.~(\ref{transofm}) and
\begin{eqnarray}&&
\int \! \cos\phi Y_{lm}d\phi= \sqrt{\frac{(2l+1)\pi }{4 l(l+1)}}P_l^1(\cos{\theta})\left(\delta_{m1}-\delta_{m, -1}\right).\nonumber
\end{eqnarray}
we further obtain that
\begin{eqnarray}&&
C^H\!=\! \lambda h^2 e^{\lambda}\sum_{l }\sqrt{\frac{(2l\!+\!1)\pi }{4 l(l\!+\!1)}}\int_{\pi/2}^{\pi}\!\frac{\tan^2{\theta}}{\cos{\theta}}\re^{- \lambda r}
({\tilde c}^{r}_{l1}\!-\!{\tilde c}^{r}_{l, -1})
\nonumber\\&&
\times\left[\frac{4l^2\!-\!1}{2\lambda r} \kappa_{l-1}\left(\frac{1}{\lambda r}\right)\!+\!
\kappa_{l-2}\left(\frac{1}{\lambda r}\right) \right]_{r\!=\!\frac{-h}{\cos{\theta}}} P_l^1(\cos{\theta}) d\theta. \nonumber
\end{eqnarray}
The exponential factor confines the integration domain to small vicinity of $\theta\!=\!\pi$. Performing expansion similar to that in  Eq.~(\ref{ian}) it follows that
\begin{eqnarray}&&
C^H\!=\!- e^{\lambda(1-h)} \sum_{l} \lambda \left({\tilde c}^{r}_{l1}-{\tilde c}^r_{l, -1}\right) \sqrt{\frac{(2l+1)\pi }{l(l+1)}} f_l(-1)
\nonumber\\&&
\times\frac{1}{\lambda^2}  \left(\frac{4l^2\!-\!1}{2\lambda h}\!+\!1\right). \label{ch}
\end{eqnarray}
We next consider the expression ${\tilde c}^{r}_{l1}\!-\!{\tilde c}^r_{l, -1}$. It follows from Eqs.~(\ref{coefficients}) that
\begin{eqnarray}&&
{\tilde c}^{r}_{l1}\!-\!{\tilde c}^r_{l, -1}\!=\!\frac{\sqrt{2l\!+\!1}\int \! P_l^1(\cos{\theta})\cos{\phi}  \left(\nabla_s\!\cdot\!\bm u \!-\!(l\!+\!2)u_r\right)d\Omega}{l(l\!+\!1)\kappa_{l-1}(\lambda^{-1})\sqrt{\pi l(l+1)} },
\nonumber
\end{eqnarray}
where we used Eq.~(\ref{sd}). It is readily seen that by using Eq.~(\ref{sp})
\begin{eqnarray}&&\!\!\!\!\!\!\!
\nabla_s\cdot(\re^{-\lambda z}\bm {\hat x})=\lambda \re^{-\lambda \cos{\theta}} \cos{\theta}\sin{\theta}\cos{\phi}, \nonumber
\end{eqnarray}
Thus it can be concluded that
\[
{\tilde c}^{r}_{l1}\!-\!{\tilde c}^r_{l, -1} \!=\!\frac{\sqrt{\pi(2l\!+\!1)}\int_{-1}^{1} \! P_l^1(x) \sqrt{1\!-\!x^2} \re^{-\lambda x} \left(\lambda x\!-\!l\!-\!2\right)dx}{l(l\!+\!1)\kappa_{l-1}(\lambda^{-1})\sqrt{ l(l\!+\!1)} }.
\] \\
By differentiating Eq.~(\ref{tbl}) over $\lambda$ we find that
\begin{eqnarray}&&
\int_{-1}^1\! \re^{-\lambda x}x\sqrt{1\!-\!x^2}P_l^{1}(x)dx
\!=\!\sqrt{\frac{2\pi}{\lambda^3}}(-1)^l l(l\!+\!1)
\nonumber\\&&
\times\left(\frac{(l\!+\!2)I_{l+1/2}(\lambda)}{\lambda}-I_{l-1/2}(\lambda)\right), \nonumber
\end{eqnarray}
where we used the identity
\begin{eqnarray}&&\!\!\!\!\!\!\!\!\!\!\!\!
\frac{dI_{l+1/2}(\lambda)}{d\lambda}=I_{l-1/2}(\lambda)-\frac{(2l+1)I_{l+1/2}(\lambda)}{2\lambda}. \nonumber
\end{eqnarray}
Collecting all terms and using Eq.~(\ref{mod}) we obtain
\begin{eqnarray}&&
{\tilde c}^{r}_{l1}\!-\!{\tilde c}^r_{l, -1} \!=\!\sqrt{\frac{\pi(2l\!+\!1)}{l(l+1)}}(-1)^{l-1}\re^{-\lambda} \frac{\pi I_{l-1/2}(\lambda)}{\lambda K_{l-1/2}(\lambda)}.
\end{eqnarray}
Substituting the last result into Eq.~(\ref{ch}) and using Eq.~(\ref{flor}) we finally find that
\begin{eqnarray}&&\!\!\!\!\!
C^H\!=\!\frac{\pi^2 }{\lambda^2} \re^{-\lambda h}\sum_{l=1}^{\infty}\frac{2(2l\!+\!1)I_{l-1/2}(\lambda)}{K_{l-1/2}(\lambda) }, \label{ch2}
\end{eqnarray}
where we have assumed that the series is fast convergent so that the relevant $l$ obeys $l^2\ll \lambda h$.

\end{appendices}



\begin{thebibliography}{99}
\bibitem{note1} We use a different (plus) sign in the definition of the resistance matrix $\bm{\mathcal R}$ from that used for the steady Stokes flow; the involved variables are complex-valued and there is no advantage in following the standard convention.
\end{thebibliography}
\end{document}